\documentclass[iop]{emulateapj}
\usepackage{amsmath}
\usepackage{graphicx}

\newcommand\Msun{\; {\rm M}_{\odot}}
\newcommand\kms{\; {\rm km}\;{\rm s}^{-1}}
\newcommand\pc{\;{\rm pc}}

\newcommand\kpc{\;{\rm kpc}}

\newcommand\yr{\; {\rm yr}}
\newcommand\Myr{\;{\rm Myr}}
\newcommand\Gyr{\;{\rm Gyr}}
\newcommand\Aunit{\Msun \yr^{-1}}
\newcommand\Surf{\Msun\;{\rm pc^{-2}}}
\newcommand\fgas{f_{\rm gas}}
\newcommand\rring{R_{\rm ring}}

\shorttitle{Effects of Gas on Bar Formation and Evolution}
\shortauthors{Seo et al.}

\begin{document}

\title{Effects of Gas on Formation and Evolution of Stellar Bars
and Nuclear Rings in Disk Galaxies}
\author{Woo-Young Seo}
\affil{Department of Physics
\& Astronomy, Seoul National University, Seoul 151-742, Republic of Korea}
\affil{Department of Physics, Chungbuk National University, Cheongju 28644, Korea}

\author{Woong-Tae Kim}
\affil{Department of Physics
\& Astronomy, Seoul National University, Seoul 151-742, Republic of Korea}
\affil{Center for Theoretical Physics (CTP), Seoul National University, Seoul 151-742, Republic of Korea}
\affil{Department of Astrophysical Sciences, Princeton University, Princeton, NJ 08544, USA}

\author{SungWon Kwak} 
\affil{Korea Astronomy and Space Science Institute, Daejeon 34055, Korea}

\author{Pei-Ying Hsieh} 
\affil{Academia Sinica Institute of Astronomy and Astrophysics, P.O. Box 23-141, Taipei 10617, Taiwan, R.O.C.}

\author{Cheongho Han}
\affil{Department of Physics, Chungbuk National University, Cheongju 28644, Korea}

\author{Phil F. Hopkins}
\affil{TAPIR, Mailcode 350-17, California Institute of Technology, Pasadena, CA 91125, USA}

\email{seowy@astro.snu.ac.kr, wkim@astro.snu.ac.kr, swkwak@kasi.re.kr,
pyhsieh@asiaa.sinica.edu.tw,
cheongho@astroph.chungbuk.ac.kr,
phopkins@caltech.edu}

\begin{abstract}
We run self-consistent simulations of Milky Way-sized, isolated
disk galaxies to study formation and evolution of a stellar bar
as well as a nuclear ring in the presence of gas.
We consider two sets of models with cold or warm disks that differ in the radial velocity dispersions, and vary the gas fraction $\fgas$ by fixing the total disk mass. A bar forms earlier and more strongly in the cold disks with larger $\fgas$, while gas progressively delays the bar formation in the warm disks . The bar formation enhances a central mass concentration which in turn makes the bar decay temporarily, after which it regrows in size and
strength, eventually becoming stronger in models with smaller $\fgas$. Although all bars rotate fast in the beginning, they rapidly turn to slow rotators. In our models, only the gas-free, warm disk undergoes rapid buckling instability, while other disks thicken more gradually via vertical heating. The gas driven inward by the bar potential readily forms a star-forming nuclear ring.
The ring is very small when it first forms and grows in size over time. The ring star formation rate is episodic and bursty due to feedback, and well correlated with the mass inflow rate to the ring.  Some expanding shells produced by star formation feedback are sheared out in the bar regions and collide with dust lanes to appear as filamentary interbar spurs. The bars and nuclear rings formed in our simulations have properties similar to those in the Milky Way.
\end{abstract}


\section{Introduction\label{sec:intro}}

More than $30\%$ of disk galaxies in the local universe possess
a well-developed stellar bar (e.g., \citealt{sel93, lee12a, gav15}).
Stellar bars greatly influence evolution of gas in disks by
exerting non-axisymmetric gravitational torque and create
gaseous substructures such as dust lanes and nuclear rings
(e.g., \citealt{san76, ath92, but96, mar03a, mar03b, kim12a}).
Gas in orbital motions hits dust lanes
and loses angular momentum to infall toward the galaxy center.
The infalling gas is gathered to form a nuclear ring where
intense star formation takes place
(e.g., \citealt{bur60, phi96, but96, kna06, maz08, maz11,com10, san10, hsi11}). Some galaxies possess filamentary interbar spurs that are connected almost perpendicularly to dust lanes
(e.g., \citealt{she00,she02,zur08}), although their origin has been
unidentified so far.

To explain the formation of gaseous substructures and understand
what controls their physical properties, a number of previous studies
employed a fixed gravitational potential to represent a stellar bar
(e.g., \citealt{ath92,eng97,mac02,mac04,reg03,reg04,ann05,kim12a,kim12b,
 kimst12,li15,shi17}).
 These studies found that dust lanes are shocks \citep{ath92} lying
almost parallel to the trajectories of $x_1$-orbits in a steady state, while the shape of a nuclear ring is well described by $x_2$-orbits
(e.g., \citealt{ath92, eng97,pat00,kim12b,li15}).
Nuclear rings form by the centrifugal barrier that the infalling, rotating gas cannot overcome, rather than resonances, and are smaller in galaxies with stronger bars \citep{kim12a}, consistent with
the observations of \citet{com10}.
Although these models with fixed bar potentials are useful to explore
the parameter space in great detail,
they are unrealistic in that stellar bars in real galaxies form
and evolve so that their properties such as strength, size, and pattern speed can vary considerably with time.

There have been numerous $N$-body simulations on how stellar bars form and evolve. These numerical work found that bars form due to dynamical instabilities of self-gravitating stellar disks \citep{mil70,hoh71,kal72,gol79,com81,sel93,pol13,sah18}.
Recent pure $N$-body simulations showed that not only the bar strength and length but also the pattern speed continuously vary with time on the course of disk evolution (\citealt{min12, man14}).
Sometimes, when the vertical velocity dispersion becomes very small compared to the radial velocity dispersion,
bars can undergo vertical buckling instability which in turn makes the bars weaker and shorter \citep{com81, com90, rah91, mer94, mar06, kwa17}. The bar properties and their temporal evolution appear to be
quite sensitive to the initial galaxy models.  For instance,
\citet{sah18} very recently showed that the bar strength at late time
can differ, by more than a factor of two, depending on the
bulge mass and density structure in the initial galaxy models.
The bar growth time as well as its strength are also dependent upon
the halo spin parameter \citep{col18}.
While these results are informative, they are based on
models with no gaseous component, and thus cannot tell how gas responds to the bars and forms substructures, not to mention how star formation occurs in real barred galaxies.

In recent years, several studies adopted Smoothed Particle Hydrodynamics simulations to include the effects of the gaseous component on stellar bars
(e.g., \citealt{fux99,bou05, ber07, ath13, ren13, car16}).
Since gas is dynamically highly responsive, it can readily change
the density distribution of the whole disk to affect the bar
formation and evolution. However, the results of the studies mentioned above differ, both quantitatively and qualitatively, in the
effects of gas on the bar formation.
For instance, \citet{ber07} found no significant correlation between the gas fraction $\fgas$ and the bar formation time, while
\citet{ath13} reported that disks with larger $\fgas$ stay
longer in a near-axisymmetric state and form a bar more slowly.
On the other hand,
\citet{rob17} showed that disks with larger $\fgas$ form bars earlier
when feedback from active galactic nuclei (AGN) is considered, while
the bar formation without AGN feedback is almost
independent of $\fgas$.

The presence of gas which is dissipative in nature appears to make stellar bars weakened or destroyed to some extent. \citet{bou05} argued that gas can completely dissolve a bar within $\sim2\Gyr$ by exerting gravitational torque, while \citet{ber07} and \citet{ath13} found that bars are not completely destroyed even in the presence of the gaseous component. In particular, \citet{ber07} found that the bar weakening in gas-poor disks is caused by buckling instability, whereas a central mass concentration (CMC) due to gas infall and pile-up near the galaxy center in gas-rich galaxies heats the disks and thus weakens the bars. They further showed that the bar strength after the weakening does not differ much in disks with different $\fgas$.
\citet{ath13} showed that bars, albeit not completely destroyed, weaken more strongly in galaxies with larger $\fgas$.
These results suggest that the role of gas on the evolution of
stellar bars is not yet clearly understood.

In this paper, we run high-resolution simulations of Milky Way-sized, isolated disk galaxies consisting of a live halo, a stellar disk, and a gaseous disk. These three components interact with each other through mutual gravity, while the gaseous component suffers radiative heating and cooling and is subject to star formation and related feedback. \citet{fux99} ran similar simulations
specific to the Milky Way to model the Galactic bar and
the central molecular zone (CMZ), but did not allow
for star formation and ensuing feedback.
Our main objectives are two-folds. First, we want to understand how the gaseous disk affects the formation and evolution of a stellar bar. Second, we want to study how a nuclear ring evolves under the situation where the bar properties vary self-consistently with time. The high-resolution models presented in this work improve the previous simulations mentioned above that did not have sufficient resolution to investigate gaseous structures and star formation in detail. These models also extend the previous hydrodynamic simulations with fixed bar potentials by allowing stellar bars to evolve over time. To explore how bars and nuclear rings develop in various situations, we vary the velocity anisotropy parameter (or Toomre stability parameter) as well as the gas fraction while keeping the total disk mass fixed.

The remainder of the paper is organized as follows. In Section \ref{sec:method}, we describe our galaxy models and numerical methods that we adopt. In Section \ref{sec:bar}, we present how stellar bars form and evolve in the presence of the gaseous component.  In Section \ref{sec:gas}, we describe evolution of gaseous structures that form and star formation rates in the nuclear, bar, and outer disk regions. In Section \ref{sec:sum_dis}, we summarize and discuss the astronomical implications of the
present work.

\section{Models and Methods}\label{sec:method}

\subsection{Galaxy Models}\label{sec:model}

To study formation and evolution of a stellar bar and its gravitational interactions with the gaseous component, we consider galaxy models with physical properties similar to those of the Milky Way. Our initial galaxy models consist of a stellar disk, a gaseous disk, a dark matter (DM) halo, and a central supermassive black hole.
For the density distribution of the DM halo, we adopt the \citet{her90} profile
\begin{equation}
\rho_{\rm DM}(r) = \frac{M_{\rm DM}}{2\pi}\frac{r_h}{r(r+r_h)^3}\;,
\end{equation}
where $r$ is the radial distance, and
$M_{\rm DM}$ and $r_h$ denote the total mass and the scale radius of the halo, respectively.  The scale radius is often specified in terms of the concentration parameter $c$ and the virial radius $r_{200}$
through
\begin{equation}
r_h = \frac{r_{200}}{c}\left[ 2\ln(1+c)-\frac{2c}{1+c}\;\right],
\end{equation}
(\citealt{spr05}).
For all models, we fix $M_{\rm DM}= 3.1\times 10^{11} \Msun$,
$c=24$, and $r_{200}=110\kpc$, corresponding to $r_h=10.7\kpc$. Initially,
we place a supermassive black with mass $M_{\rm BH} = 4\times 10^6\Msun$ at the galaxy center, which is allowed to accrete surrounding gas without any feedback effect in the present work.

For the stellar disk, we initially adopt the following density distribution
\begin{equation}\label{e:rhos}
\rho_s(R,z)=\frac{M_s}{4\pi z_s R_s}\exp\left(-\frac{R}{R_s}\right)
{\rm sech}^2 \left(\frac{z}{z_s}\right),
\end{equation}
where $R$ and $z$ are the radial and vertical distances in the cylindrical coordinates, while $M_s$, $R_s=3\kpc$, and $z_s=0.3\kpc$ refer to the mass, the radial scale length, and the vertical scale height of the stellar disk, respectively.  For models with gas included, we adopt the same form as Equation \eqref{e:rhos} for the initial density distribution $\rho_g$ of a gaseous disk, but with the gas mass $M_g$, the scale radius $R_g=3\kpc$, and the scale height $z_g=0.1\kpc$. To study the effect of gas on the bar formation, we vary the gas fraction $\fgas \equiv M_g/(M_g+M_s)$ in the range of $0$ to $10\%$, while fixing the total disk mass to $M_{\rm disk} = M_s + M_g = 5 \times 10^{10} \Msun$.
The observed range of the gas fraction for Milky Way-like galaxies
is around $\sim1$--$10$\% (e.g., \citealt{pap16}).

\begin{figure}
\epsscale{1.1} \plotone{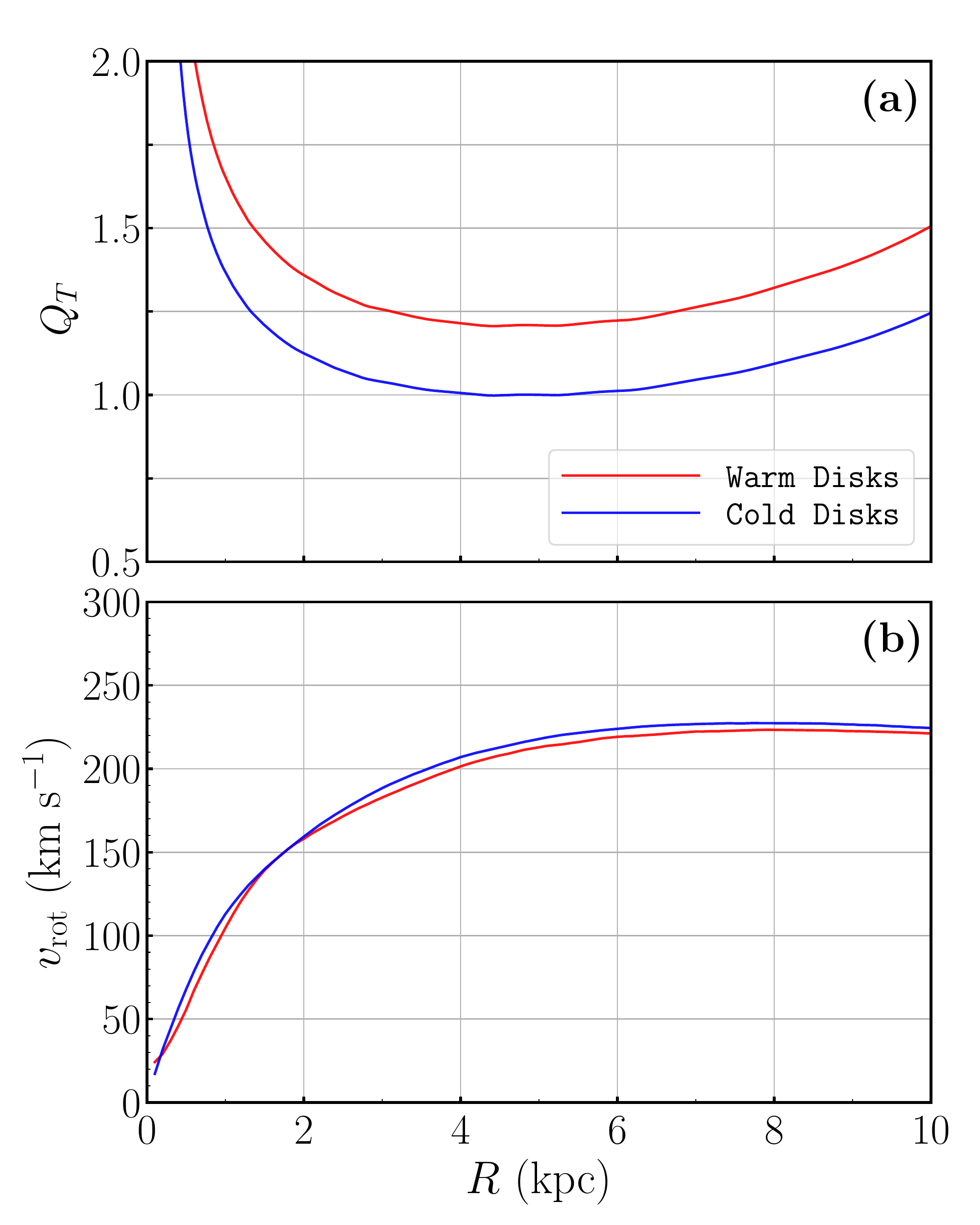}
\caption{
Radial distributions of (a) the Toomre stability parameter
$Q_T$ and (b) rotational velocity $v_{\rm rot}$
in models with a cold disk (blue) or a warm disk (red).
\label{fig:tq}}
\end{figure}

To construct a stellar disk by distributing particles, one needs to specify the velocity anisotropy parameter
\begin{equation}\label{e:fR}
  f_R = \frac{\sigma_R^2}{\sigma_z^2},
\end{equation}
where $\sigma_R$ and $\sigma_z$ are the velocity dispersions in the radial and vertical directions, respectively (\citealt{yu14}).
For fixed $\sigma_z$, varying $f_R$ corresponds to changing \citet{too66} stability parameter
\begin{equation}\label{e:Too}
  Q_T=\frac{\kappa\sigma_R}{3.36G\Sigma_{\rm disk}}=f_R^{1/2}\frac{\kappa\sigma_z}{3.36G\Sigma_{\rm disk}},
\end{equation}
where $\kappa$ is the epicycle frequency and $\Sigma_{\rm disk} = \Sigma_s + \Sigma_g = \int(\rho_s+\rho_g)dz$ is the surface density of the combined (stellar plus gaseous) disk.  To study the effect of the velocity anisotropy (or $Q_T$) on the bar formation, we in this paper consider two sets of disk models: relatively cold disks with $f_R=1.0$ and relatively warm disks with $f_R=1.44$. These initial values of $f_R$ without a bar are smaller than the observed values $f_R\sim4$ for the Milky Way in the solar neighborhood (e.g., \citealt{sha14,gui15,kat18}): we will show in Section \ref{sec:buck} that the bar formation and evolution increase $f_R$ in
our models to the values close to 4. Figure \ref{fig:tq}(a)
plots the radial distributions of $Q_T$ for the cold and warm disks.
Note that $Q_T$ is minimized at $R\approx 5\kpc$, with the minimum values of
$Q_{T, \rm min}=1.0$ for the cold disks  and
$Q_{T, \rm min}=1.2$ for the warm disks. Table \ref{tbl:model} lists the names and parameters of all models together with some numerical outcomes. The models with postfix ``{\tt C}" and ``{\tt W}" have a cold and warm disk, respectively, and the number after the postfix represents the gas fraction $\fgas$ in each set of models.

\begin{deluxetable*}{ccccccccc}
\tablecaption{Model Parameters and Simulation Outcomes\label{tbl:model}}
\tablenum{1}
\tablehead{\colhead{Model} & \colhead{$\fgas$} & \colhead{$f_R$} & \colhead{$\langle A_2\rangle$} & \colhead{$\langle\mathcal{R}\rangle$} & \colhead{$\langle R_{\rm ring}\rangle$}& \colhead{$\langle \text{SFR}_{\rm ring}\rangle$} \\
(1) & (2) & (3) & (4) & (5) & (6) & (7)
}
\startdata
{\tt C00}  &   -  & 1.0 & 0.63 & 1.67 & -    & -    \\
{\tt C05}  & $5$  & 1.0 & 0.49 & 1.63 & 0.40 & 0.19  \\
{\tt C07}  & $7$  & 1.0 & 0.47 & 1.83 & 0.35 & 0.13  \\
{\tt C10}  & $10$ & 1.0 & 0.22 & 2.50 & 0.50 & 0.04  \\
\hline
{\tt W00}  &   -  & 1.44& 0.59 & 1.51 & -    & -    \\
{\tt W05}  & $5$  & 1.44& 0.52 & 1.60 & 0.27 & 0.20 \\
{\tt W07}  & $7$  & 1.44& 0.46 & 1.51 & 0.29 & 0.31  \\
{\tt W10}  & $10$ & 1.44& 0.36 & 1.60 & 0.19 & 0.19  \\
\enddata
\tablenotetext{}{Note. The brackets $\langle\,\rangle$ denote the late-time temporal
average over $t=4.5$--$5.0\Gyr$. Column 1: model name.
Column 2: initial gas fraction (\%).
Column 3: initial velocity anisotropy parameter (Equation \ref{e:fR}).
Column 4: time-averaged bar strength;
Column 5: time-averaged ratio of the corotation radius to the bar length.
Column 6: time-averaged nuclear ring size (kpc).
Column 7: time-averaged SFR in the ring ($\rm M_\odot\,yr^{-1}$).}
\end{deluxetable*}

Our initial galaxy models with no gas are realized by making use of the publicly available GALIC code \citep{yu14}.
GALIC is very flexible in generating an equilibrium configuration. It adjusts particle velocities iteratively to obtain a desired density distribution. For models with gas, we reduce the mass $m_s$ of each stellar particle to $m_s (1-\fgas)$, while keeping their number and positions intact. We then insert a self-gravitating, isothermal gas disk with mass $\fgas M_{\rm tot}$ and the vertical scale height $z_g=0.1\kpc$.
Since the conversion of a part of the stellar disk to the self-gravitating gaseous disk effectively reduces the scale height and velocity dispersions, the new hybrid disk is slightly out of equilibrium.
We thus evolve the whole system over $0.1\Gyr$ by imposing an isothermal condition and no star formation.  The system gradually relaxes to a quasi-equilibrium state in which the stellar disk
remains almost unchanged with $z_s\simeq 0.3\kpc$, while the gaseous disk, being more dynamically responsive, becomes thinner to $z_g\approx50$ and $80\pc$ at $R=3$ and $5\kpc$, respectively.
While $z_g$ at the relaxed state tends to be smaller for larger $\fgas$ and smaller $f_R$, the differences are only within a few percents. Figure \ref{fig:tq}(b) plots the rotational velocities $v_{\rm rot}$ at the relaxed state for the cold- and warm-disk models, insensitive to $\fgas$.  This indicates that the total gravitational potential in the hybrid disk is almost unchanged notwithstanding the gas fraction.

Each model is constructed by distributing a total of $1.1\times10^7$ particles: $N_h=5\times10^6$, $N_s=5\times10^6$, and $N_g=1\times10^6$ for the halo, stellar disk, and gaseous disk, respectively. The mass of each halo particle is $m_h=6.2\times10^4\Msun$, while stellar and gaseous particles each have mass of $m_s = 9.5\times10^3\Msun$ and
$m_g=2.5\times10^3\Msun$ for models with $\fgas=5\%$,
and $m_s=9.0\times10^3\Msun$ and $m_{g}=5.0\times10^3\Msun$ for models with $\fgas=10\%$.

\subsection{Numerical Method}

We evolve our galaxy models using the GIZMO code \citep{hop15},
which is a second-order accurate, magnetohydrodynamics code based on a kernel discretization of the volume coupled to a high-order matrix gradient estimator and a Riemann solver. It thus conserves mass, momentum, and energy almost exactly. Gravity is solved by an improved version of the Tree-PM method with an opening angle of $\theta=0.7$.
Softening lengths for stellar and halo particles are set to $10\pc$ and $50\pc$, respectively, corresponding to the mean particle separations. The gaseous particles have a fully adaptive softening
length with the minimum value set to $1\pc$.

For the evolution of gaseous particles, we use a second-order accurate meshless finite-mass method that conserves angular momentum
very accurately \citep{hop15}. The gaseous particles are subject to radiative cooling due to various line emissions
and heating mostly via photoionization and photoelectric effects
(e.g., \citealt{kat96,hop14,hop18}). We find that these radiative processes together with star formation feedback and bar-related gas dynamics result in $T\sim10^4\rm\,K$ for the temperature of most gas in the disk, corresponding to the warm neutral gas of the interstellar medium.

We implement a stochastic prescription for star formation and feedback. Star formation is allowed to occur only in dense, self-gravitating regions where the velocity field is converging and
the local gas density exceeds the critical value
$n_{\rm crit}=10\;{\rm cm}^{-3}$.
For a gaseous particle satisfying the above criteria,
the star formation probability over the time interval $\Delta t$ is
given by $p=1-\exp(-\epsilon_{\rm ff}\Delta t/t_{\rm ff})$, where $\epsilon_{\rm ff}\approx 1\%$ is the star formation efficiency
\citep{hop11,seo13}. In each time step, we generate a uniform random number $\mathcal{N}\in[0,1)$ and switch such a gaseous particle into a new stellar particle with the same mass only when $\mathcal{N}>p$.

We handle star formation feedback using simple momentum input
as well as mass return to the neighboring gaseous particles
in the form of Type Ia and II supernova (SN) events.
In our models, each stellar particle (with mass approximately $\sim10^4\Msun$) corresponds to an unresolved star cluster.
Assuming the \citet{kro01} initial mass function and using
the lifetime of Type II SN progenitors (\citealt{lej01}) and
the rate of Type Ia SNe (\citealt{man06}), we calculate the number of SN events, $N_{\rm SN}$, expected from a stellar particle in each time step. Type II SNe occur only from newly formed particles younger than $10\,\Myr$, the lifetime of $8\Msun$ stars, while  Type Ia SNe can explode from not only newly-created stars older than $10\,\Myr$ but also the pre-existing particles
comprising of the initial stellar disk. SNe inject momentum and mass to the surrounding gas particles inside the shell radius
\begin{equation}
r_{\rm sh} = 0.025 \;N_{\rm SN}^{1/4}\;
  \left(\frac{n}{1\, {\rm cm}^{-3}}\right)^{-1/2}\kpc,
\end{equation}
corresponding to the shock radius at the shell formation stage for
$N_{\rm SN}$ almost-simultaneous SNe.
The amount of the total radial momentum deposited is given by
\begin{equation}
P_{\rm SN} = 2.8\times10^5 \;N_{\rm SN}^{7/8}\;\left(\frac{n}{0.1 {\rm cm}^{-3}}\right)^{-0.17} \Msun \kms
\end{equation}
(e.g., \citealt{che74,shu80,cio88,seo13,kim15}).
Each gas particle inside $r_{\rm sh}$
receives momentum and mass proportional to its occupied volume.

\begin{figure*}
\epsscale{1.2} \plotone{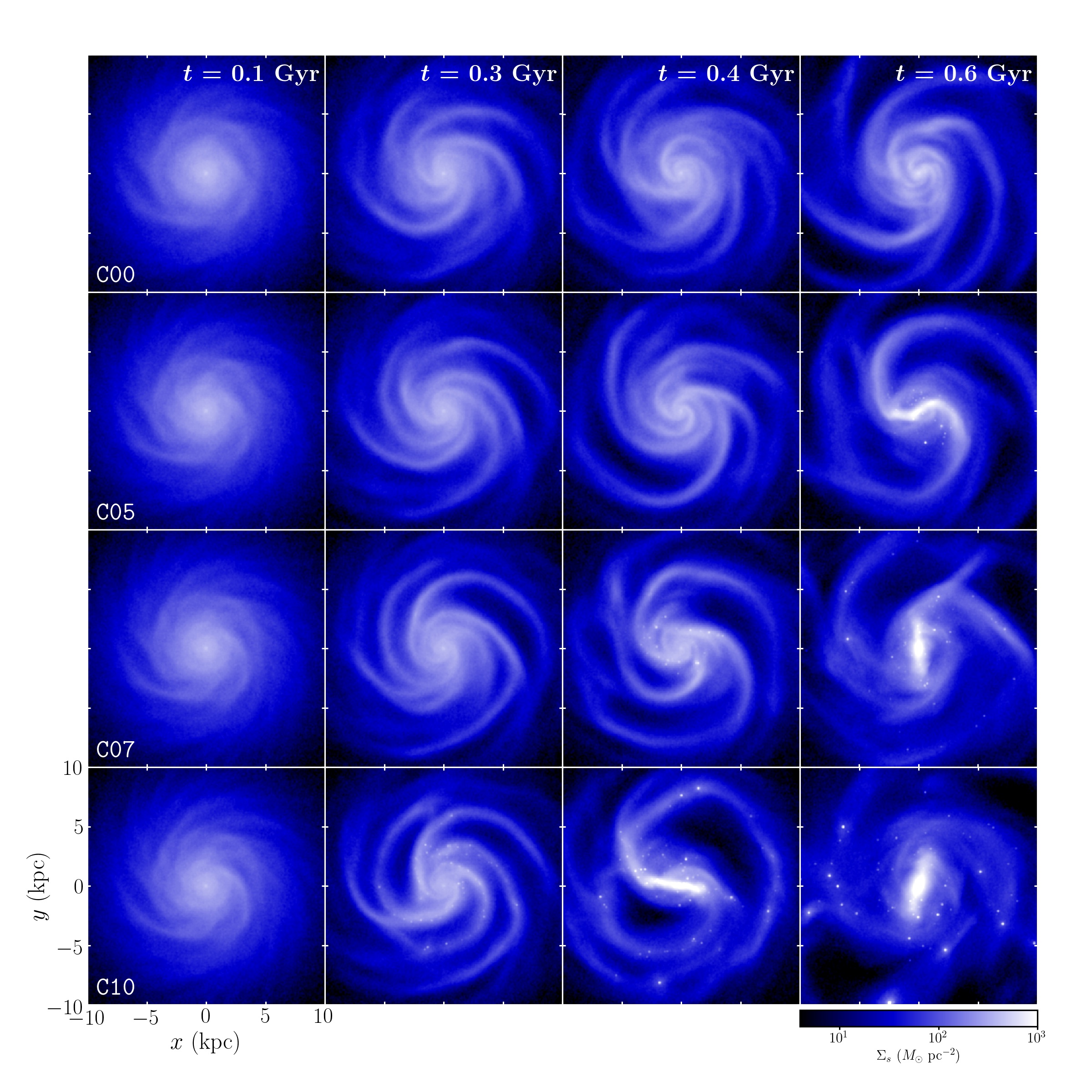}
\caption{
Snapshots of logarithm of the stellar surface density $\Sigma_s$
for the cold-disk models at $t=0.1$, $0.3$, $0.4$, and $0.6\Gyr$ from left to right. Each row corresponds to model with $\fgas=0$, 5, 7, and 10\% from top to bottom. Discrete bright spots represent newly formed stars from the gas disk, while smooth color distributions display the stellar particles in the initial disk. Colorbar labels $\Sigma_s/(\rm M_\odot \pc^{-2})$.}
\label{fig:snap_u}
\end{figure*}

\begin{figure*}
\epsscale{1.0}
\plotone{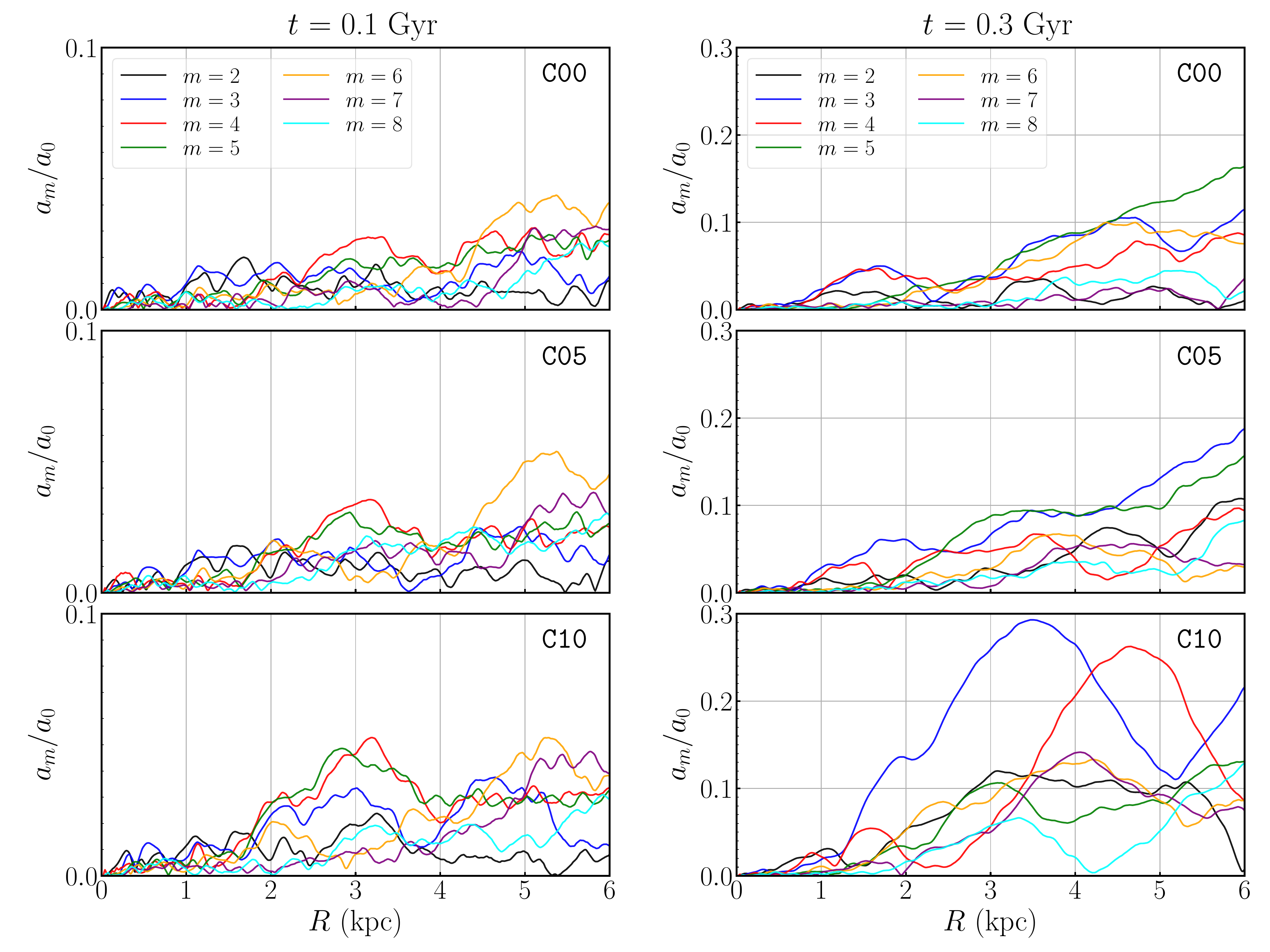}
\caption{
Radial distributions of the normalized Fourier amplitudes $a_m/a_0$ of the stellar surface density for models
{\tt C00}, {\tt C05}, and {\tt C10} at $t=0.1\Gyr$ (left) and $t=0.3\Gyr$ (right).
\label{fig:am_cold}}
\end{figure*}

\section{Stellar Bars}\label{sec:bar}

In this section, we focus on the formation and evolution of stellar bars and the effects of the gaseous component on them. Evolution of gaseous structures including nuclear rings and star formation therein
will be presented in Section \ref{sec:gas}.

\subsection{Bar Formation}\label{sec:early}

Since finite disk thickness reduces self-gravity at the disk midplane, all combined disks with $Q_{T,\rm min}\geq1$
are stable to \emph{axisymmetric} gravitational perturbations.
However, \emph{non-axisymmetric} perturbations are still able to grow as they swing from leading to trailing configurations (e.g., \citealt{bin07,kim07,kwa17}), eventually organizing into bars.
We find that the effect of gas on bar formation is different between models with a cold disk and a warm disk, as described below.

\subsubsection{Cold-disk Models}\label{sec:cold}

Figure \ref{fig:snap_u} plots the stellar surface density in logarithmic scale in the $10\kpc$ regions of the cold-disk models at $t=0.1$, 0.3, 0.4, and $0.6\Gyr$.
It is apparent that the disks at early time are subject to swing amplification and produce spiral structures
that extend from the galaxy center all the way to the outer edge.
The disks harbor various spiral modes with high azimuthal
mode numbers ($m=3$--$6$), with their amplitudes depending on $\fgas$. Model {\tt C10} has strongest spiral arms since its disk is effectively coldest, and newly formed stars indicated by bright spots are distributed along the spirals at $t=0.3\Gyr$.

Figure \ref{fig:am_cold} plots the radial distributions of the various Fourier amplitudes $a_m$ relative to $a_0$ in the stellar disks of models {\tt C00}, {\tt C05}, and {\tt C10} at $t=0.1$ and $0.3\Gyr$.  At $t=0.1\Gyr$, the modes with $m=3,4$ and $m=6,7$
have largest amplitudes in the $R\sim2$--$4\kpc$ and $R\sim4$--$6\kpc$ regions of the disks, respectively.
The strength of swing amplification is measured roughly by
the instantaneous growth rate multiplied by the duration of amplification (e.g., \citealt{jul66,kim01}).
While modes with high $m$ may have a large instantaneous growth rate, they usually have a limited time for amplification because they quickly wind out kinematically due to background shear (e.g., \citealt{gol65,jul66}).  It turns out that the $m=3$ mode grows most strongly at $R\lesssim 3\kpc$ in all the cold-disk models, although modes with $m=4$ and $5$ also contribute to the perturbed density in the outer regions.
The dominance of the three-arm spiral modes at small radii
in the early phase of bar formation is also seen in the $Q_{T,\rm min}=1$ models of \citet{fan15}.

\begin{figure*}
\epsscale{1.2} \plotone{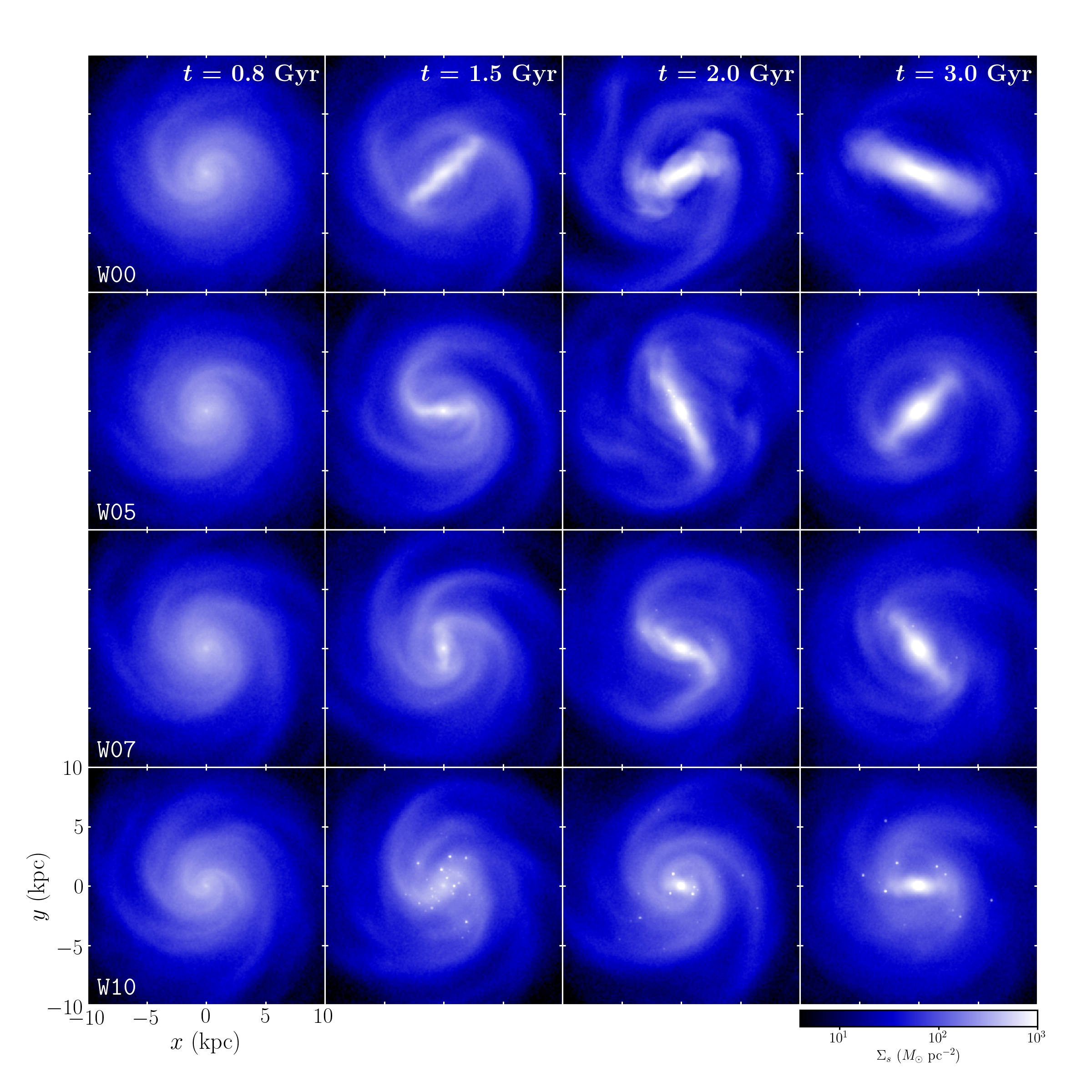}
\caption{Snapshots of logarithm of the  stellar surface density
$\Sigma_s$ for the warm-disk models at $t=0.8$, $1.5$, $2.0$, and $3.0\Gyr$ from left to right.  Each row corresponds to model with $\fgas = 0$, 5, 7, and 10\% from top to bottom.  Colorbar labels $\Sigma_s/(\rm M_\odot \pc^{-2})$.
\label{fig:snap_s}}
\end{figure*}

\begin{figure}
\epsscale{1.0} \plotone{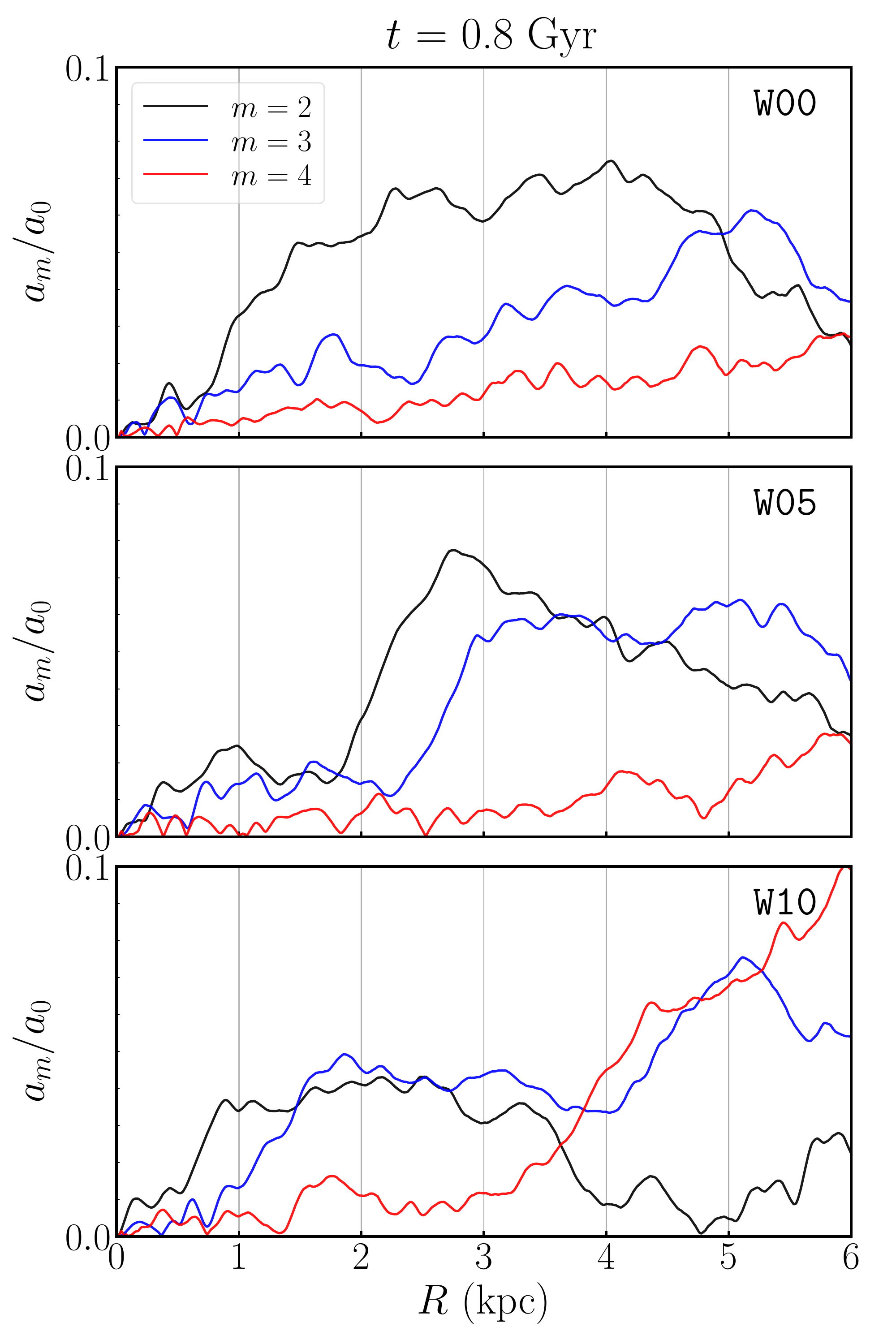}
\caption{Radial distributions of the normalized Fourier amplitudes $a_m/a_0$ of the stellar surface density for $m=2$ to $4$ modes in the models {\tt W00}, {\tt W05}, and {\tt W10} at $t=0.8\Gyr$.
\label{fig:am_warm}}
\end{figure}

Swing amplification in the cold-disk models is so virulent that the spirals rapidly become nonlinear. For instance, the $m=3$ spirals in model {\tt C10} at $t=0.3\Gyr$ have an amplitude $\delta\Sigma_s/ \Sigma_s\sim 0.5$ at $R\sim1$--$3\kpc$.
These spirals interact nonlinearly with other spirals with higher
$m$ that propagate radially inward.  As a consequence,
one arm of the $m=3$ spirals becomes loose and merge with the other two arms, eventually transforming into an $m=2$ bar mode
that is supported by stable $x_1$-orbit families \citep{con89}.
Since gaseous particles are colder than stellar particles,
the swing amplification and ensuing bar formation occur faster in models with larger $\fgas$. Figure \ref{fig:snap_u} shows that model {\tt C10} already possesses a well-developed bar by $t=0.4\Gyr$, while it takes model {\tt C00} about $\sim0.5\Gyr$ longer to
form a noticeable bar.

\subsubsection{Warm-disk Models}

Warm disks form a bar slower than cold disks due to lager $Q_T$.
Unlike in the cold disks, the presence of gas delays the bar formation in the warm disks. Figure \ref{fig:snap_s} plots the stellar surface density in logarithmic scale in the $10\kpc$ regions of the warm-disk models at $t=0.8$, 1.5, 2.0, and $3.0\Gyr$.
Figure \ref{fig:am_warm} plots the radial distributions of the Fourier amplitudes of the $m=2$--$4$ azimuthal modes that dominate in models {\tt W00}, {\tt W05}, and {\tt W10}
at $t=0.8\Gyr$. At this time, by which all cold disks have already formed a bar, the warm disks still exhibit only weak spiral structures. Similarly to the cold disks, the warm disks are subject to swing amplification, but the associated amplification factor is less than 10\% and the resulting spiral waves after the initial swing amplification are thus in the linear regime.

Still, a warm disk with larger $\fgas$ is effectively colder and thus more vulnerable, especially at $R\sim3$--$6\kpc$ where $Q_T$ is smallest. Since random velocity dispersions (or acoustic waves) tend to stabilize small-scale modes, disks with larger $\fgas$ should favor larger-$m$ modes for growth. This expectation is consistent with Figure \ref{fig:am_warm} that shows that the modes with $m=2$, 3, and 4 have grown most strongly by $t=0.8\Gyr$ in models {\tt W00}, {\tt W05}, and {\tt W10}, respectively. This in turn indicates that the amplitude of the $m=2$ mode that will seed the bar formation
is larger in models with smaller $\fgas$. Note that these mode numbers favored in the warm disks are smaller than those dominant in the early phase of swing amplification in the cold disks.

\begin{figure*}
\epsscale{1.0} \plotone{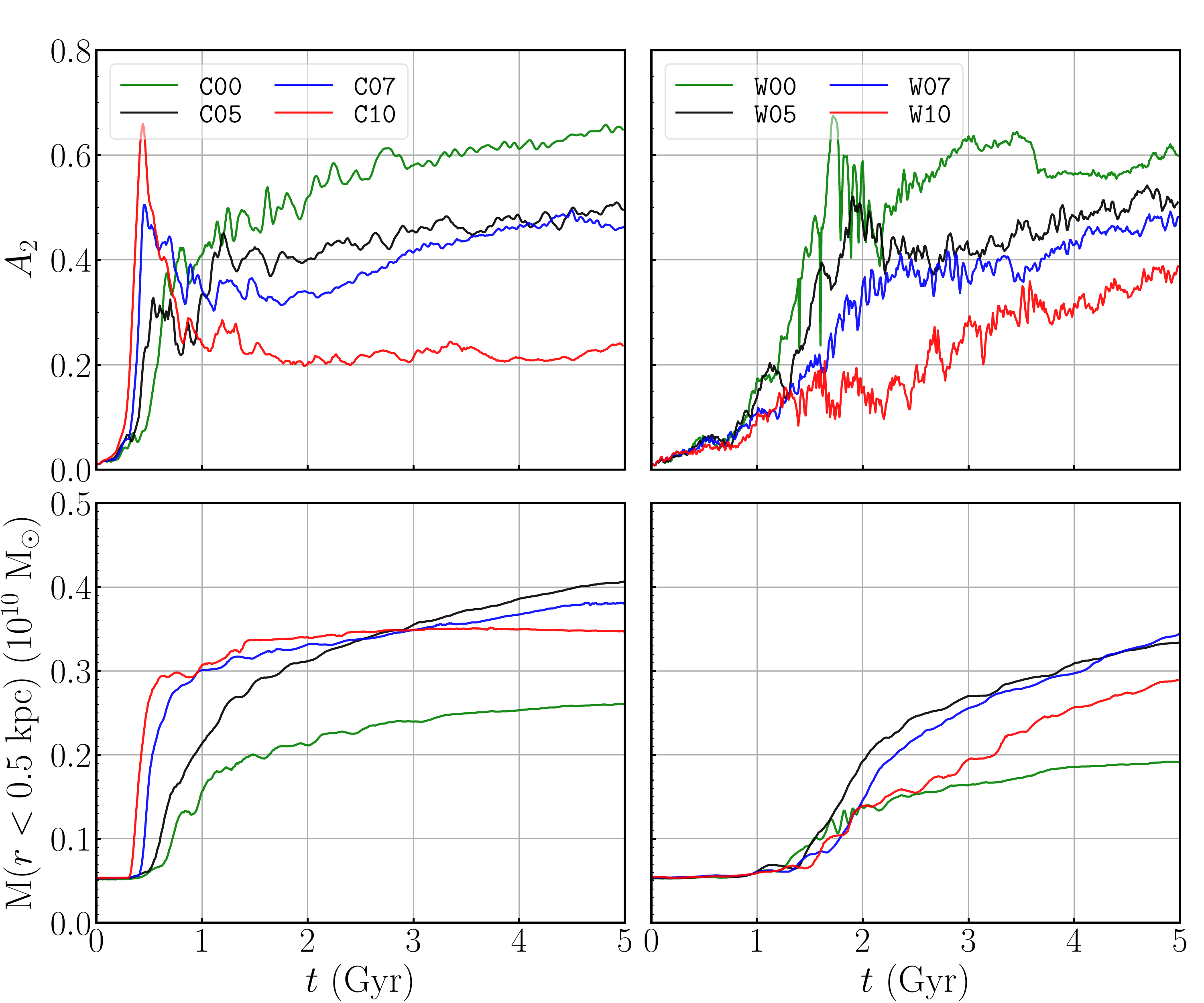}
\caption{
Temporal variations of the bar strength $A_2$ (upper panels) and
the CMC within the $r=0.5\kpc$ regions (lower panels) for the cold-disk models (left) and the warm-disk models (right).
In the cold-disk models, the bar strength increases rapidly and then decays after reaching a peak due to the increase in the CMC. In the warm-disk models, the bar and CMC grow more slowly and steadily. The bar strength at late time is anti-correlated with $\fgas$ in both cold-disk and warm-disk models.
\label{fig:str_cmc}}
\end{figure*}

Because of relatively large $Q_T$, the initial swing amplification
in the warm disks is too mild to form a bar instantly.
Without a bulge, these trailing waves are well positioned to
propagate right through the center and then emerge as leading waves
in the opposite side,
amplifying further as they unwind again from leading to
trailing configurations (e.g., \citealt{bin07}).
An eventual bar formation requires several cycles of swing amplifications and feedback loops, which takes longer than $\sim1\Gyr$. A disk with larger $\fgas$ takes longer to form a bar since the bar-forming $m=2$ perturbations are weaker
at the end of the initial swing amplification.

\subsection{Physical Properties}

\subsubsection{Bar Strength}

\begin{figure*}
\epsscale{1.0} \plotone{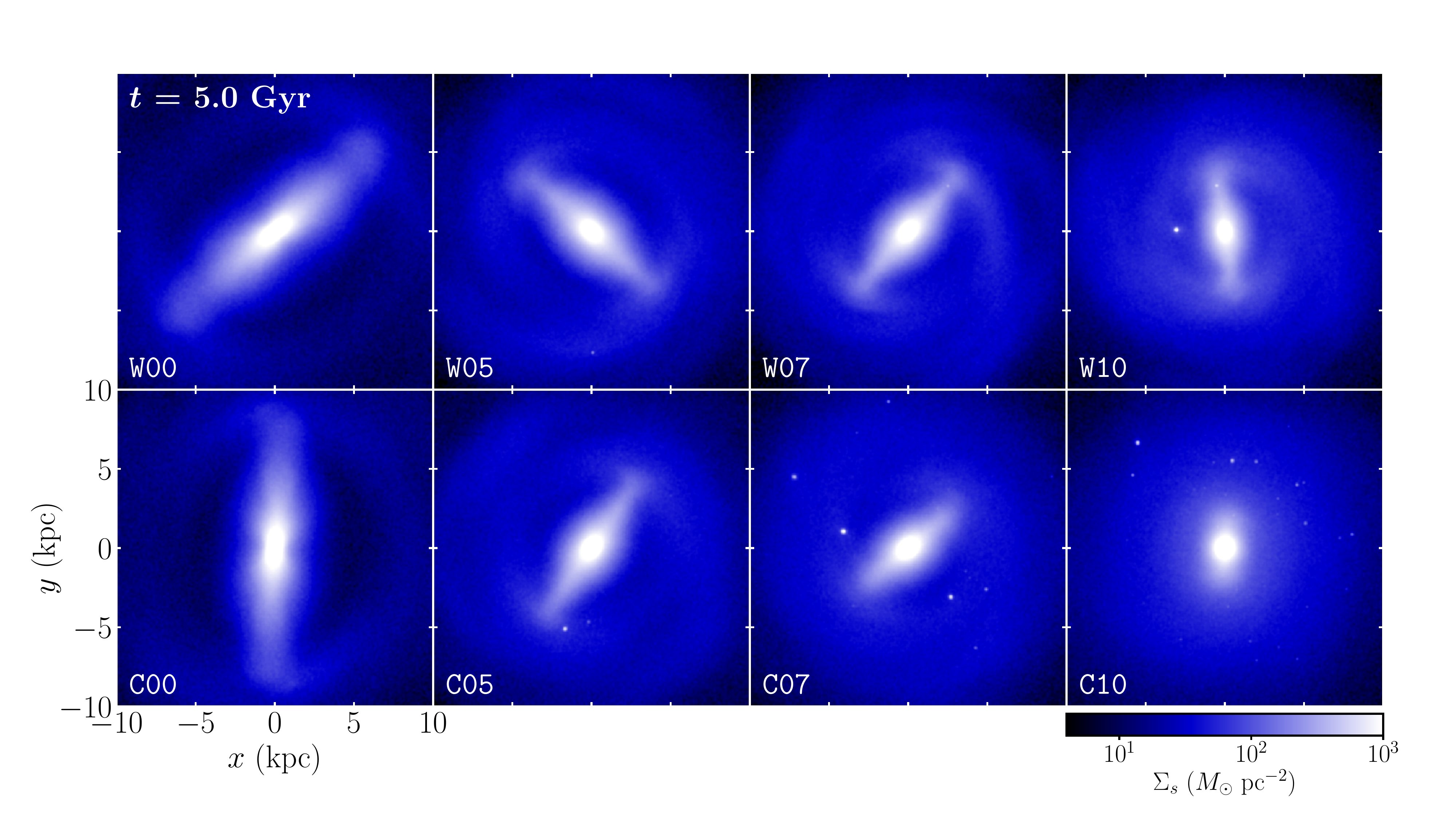}
\caption{Snapshots of logarithm of the stellar surface density
in the $10\kpc$ regions for all models at $t=5\Gyr$.
Colorbar labels $\Sigma_s/(\rm M_\odot \pc^{-2})$.
\label{fig:snap_end}}
\end{figure*}

In our models, the strength and size of bars vary with time considerably, which is related to the CMC.
Conventionally, the bar strength $A_2$ is defined by the maximum value of the $m=2$ Fourier mode relative to the $m=0$ mode as
\begin{equation}
  A_2 \equiv  \max\left\{\frac{a_2(R)}{a_0(R)}\right\}.
\end{equation}
To measure the CMC, we use the total (star plus gas) mass inside the central regions with $r\leq0.5\kpc$.
Figure \ref{fig:str_cmc} plots the temporal changes of the bar strength and the CMC for the cold-disk (left) and the warm-disk (right) models. Figure \ref{fig:snap_end} plots the distribution of the stellar surface density at $t=5\Gyr$ for all models.
Clearly, a bar forms earlier in the cold disks than in the
warm disks. The presence of gas causes the bar to form faster and more strongly in the cold-disk models, while it makes the bar formation delayed in the warm-disk models.
Bar formation necessarily involves mass relocation and thus changes the CMC even in the gas-free models, although the CMC is more significant in models with more gas since the bar potential induces strong gas inflows. Changes in the central mass affect orbits of stars, notably in the vertical
direction, when they pass close to the galaxy center, leading to bar
thickening and weakening (e.g. \citealt{mar06, ber07, kwa17}).
In addition, gas that is dissipative in nature does not
follow exact $x_1$-orbits in the bar regions and thus makes the bar weaker.

In the cold-disk models, the rapid decay of the bar strength after
attaining a peak is caused by the rapid increase in the CMC.
The bar weakening in model {\tt C10} is so dramatic that it quickly turns into an oval shape, as illustrated in Figure \ref{fig:snap_end}. With a relatively slow increase of the CMC, the bar in model {\tt C00} does not experience such weakening: it keeps longer and stronger secularly, and attains a Boxy/Peanut (B/P) shape at $t=5\Gyr$, a common late-time feature of $N$-body bars (e.g., \citealt{man14}).
The central mass increases more rapidly as the bar grows faster and stronger, resulting in stronger bar weakening in the cold-disk models with larger $\fgas$.  This collectively makes a bar stronger
in disks with smaller $\fgas$ at the end of the runs.

In the warm-disk models, on the other hand, the bar growth time is relatively long and the CMC growth is accordingly quite slow.
Therefore, the bar weakening after the peak strength is not so
dramatic as in the cold disks. As a result, the bars at $t=5\Gyr$ in the warm disks are stronger for smaller $\fgas$.
Column (4) of Table \ref{tbl:model} lists the values of
$A_2$ averaged over $t=4.5$--$5.0\Gyr$.
Putting together all the results for both cold and warm disks, we
conclude that the bar strength in the late phase of the disk evolution is inversely proportional to the gas fraction, independent of $Q_T$, notwithstanding temporal evolution in the early phase.
A mild drop in $A_2$ of model {\tt W00} at $t=3.5$--$3.7\Gyr$ is
due to buckling instability, which will be discussed
in Section \ref{sec:buck}.

\subsubsection{Bar Length and Pattern Speed}

\begin{figure*}
\epsscale{0.8} \plotone{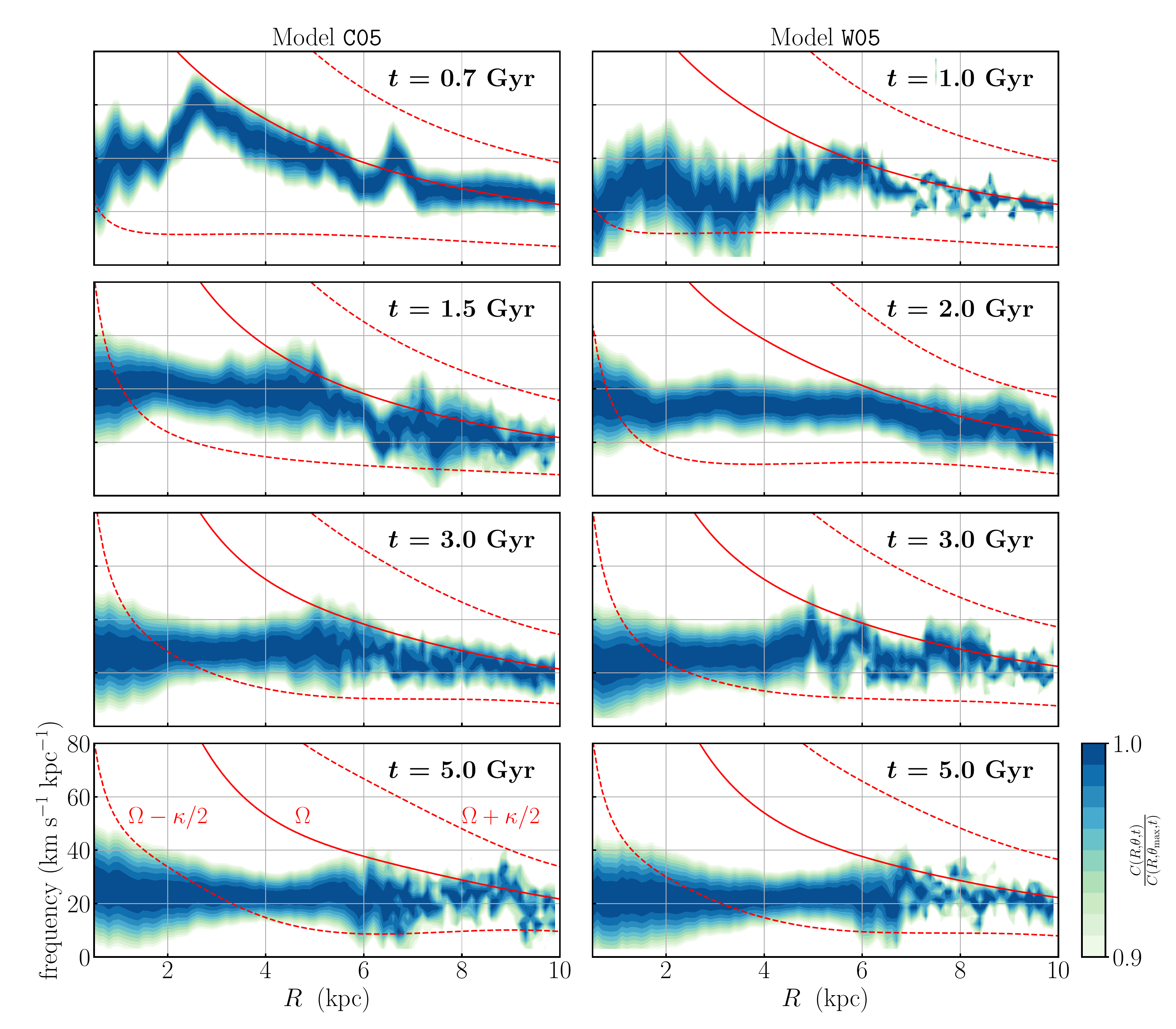}
\caption{Contours of the normalized cross-correlation of the
perturbed surface density in the radius-angular frequency plane
for models {\tt C05} (left) and {\tt W05} (right) at different epochs. The solid and dashed line in each panel plot instantaneous $\Omega(R)$ and $\Omega\pm\kappa/2$. While bars have a pattern speed $\Omega_b$ almost independent of $R$, corresponding to rigid-body rotation, the pattern speed of spirals arms is close to $\Omega$ at early time and gradually becomes similar to the bar pattern speed at late time. Colorbar labels $C(R, \theta,t)/C(R, \theta_{\rm max},t)$.}
\label{fig:ncc}
\end{figure*}

Since a bar is smoothly connected to the disk in which it is embedded, it is quite ambiguous to determine the bar ends.
We empirically find that the stellar surface density
of $\Sigma_s = 80 \Msun \,\pc^{-2}$ traces the bar boundaries reasonably well, which allows us to measure the bar semi-major axis $R_b$ from our simulations.
To calculate the pattern speeds of non-axisymmetric features, we use
the cross-correlation of the perturbed surface densities $\delta \Sigma_s \equiv  \Sigma_s -\Sigma_s(t=0)$ at two different epochs separated by $\delta t=0.1\Gyr$:
\begin{equation}\label{eq:cc}
\begin{split}
 C(R, \theta, t) \equiv &  
 \frac{1}{\Sigma_s(R,t=0)^2} \\
 & \int_0^{2\pi} \delta \Sigma_s(R,\phi,t)
               \delta \Sigma_s(R,\phi + \theta, t+\delta t) d\phi
\end{split}
\end{equation}
(e.g., \citealt{oh08,oh15}).
At a given time and radius, we determine the angle $\theta_{\rm max}$ at which $C(R, \theta, t)$ is maximized. The instantaneous pattern speed of the non-axisymmetric features is then determined by
$\Omega_p(R, t) = \theta_{\rm max}/\delta t$.

Figure \ref{fig:ncc} plots as contours the amplitudes of the normalized cross-correlation $C (R, \theta, t)/C(R, \theta_{\rm max}, t)$ in the $R$--$(\theta/\delta t)$ plane for some selected epochs of models {\tt C05} (left) and {\tt W05} (right) together with the  $\Omega$ and $\Omega\pm\kappa/2$ curves at given time.
The cross-correlation in the inner disk traces the bar pattern speed, while the regions outside the bar is dominated by the spiral arms.
The bar pattern speed varies with $R$ at very early time when
the bars are forming. Once they achieve a full strength (i.e., after $t=0.5\Gyr$ and $t=1.8\Gyr$ for models {\tt C05} and {\tt W05}, respectively), however, $\Omega_p(R, t)$ becomes almost independent of $R$, indicating that they are rotating rigidly.  On the other hand,
the spirals that form in the regions outside the bar
have pattern speeds almost equal to $\Omega$, suggesting that they are material arms, until $\sim 2\Gyr$ after the bar formation. They then tend to corotate with the bar (see Section \ref{sec:spiral}).

\begin{figure}
\epsscale{1.0} \plotone{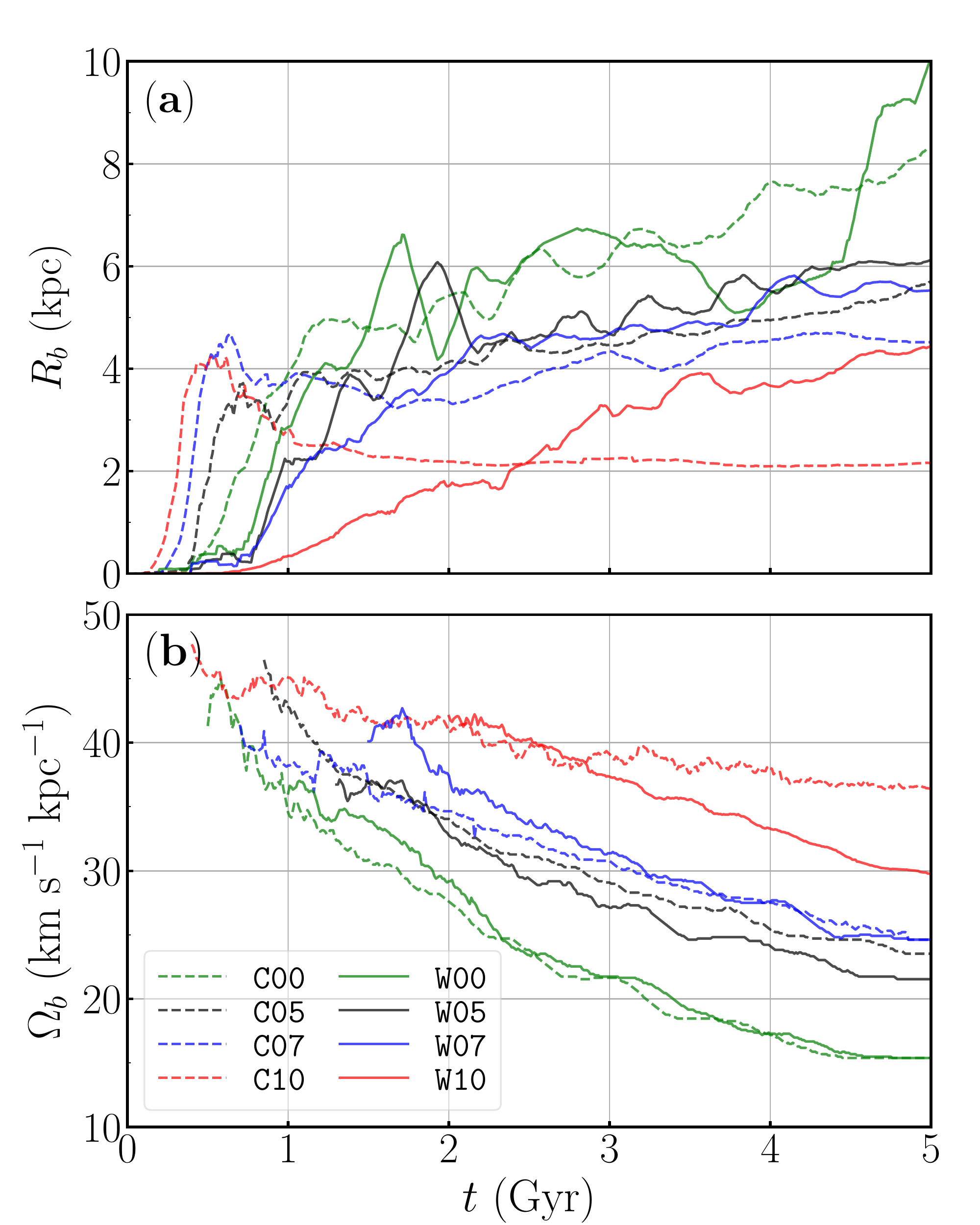}
\caption{
Temporal changes of (a) the bar semi-major axis $R_b$ and (b) the bar
pattern speed $\Omega_b=\Omega_p(R=3\kpc)$.
The solid and dotted lines correspond to the models with the warm disks and cold disks, respectively.
\label{fig:length}}
\end{figure}

Figure \ref{fig:length} plots temporal variations of $R_b$ and the bar pattern speed $\Omega_b\equiv \Omega_p(R=3\kpc)$ for all models. Overall,
the increasing and decreasing trend of $R_b$ with time is similar to
that of the bar strength, such that a bar becomes long (short) when it is strong (weak).
Bars in models {\tt C07} and {\tt C10} are
relatively short when they achieve the peak strength
(at $t\sim0.4$--$0.5\Gyr$),
indicating that their strength is largely due to high stellar density
resulting from rapid bar formation.
Note that the bar pattern speed decreases continuously after the formation in all models,
and a stronger bar slows down at a faster rate. This is because
the angular momentum transfer between the bar and halo
is more active for a stronger bar (e.g., \citealt{ath13,mar06}).
Note also that all bars except an oval in model {\tt C10} grow
in size as they slow down over time (e.g., \citealt{ath02, ath03}).

\begin{figure}
\epsscale{1.0} \plotone{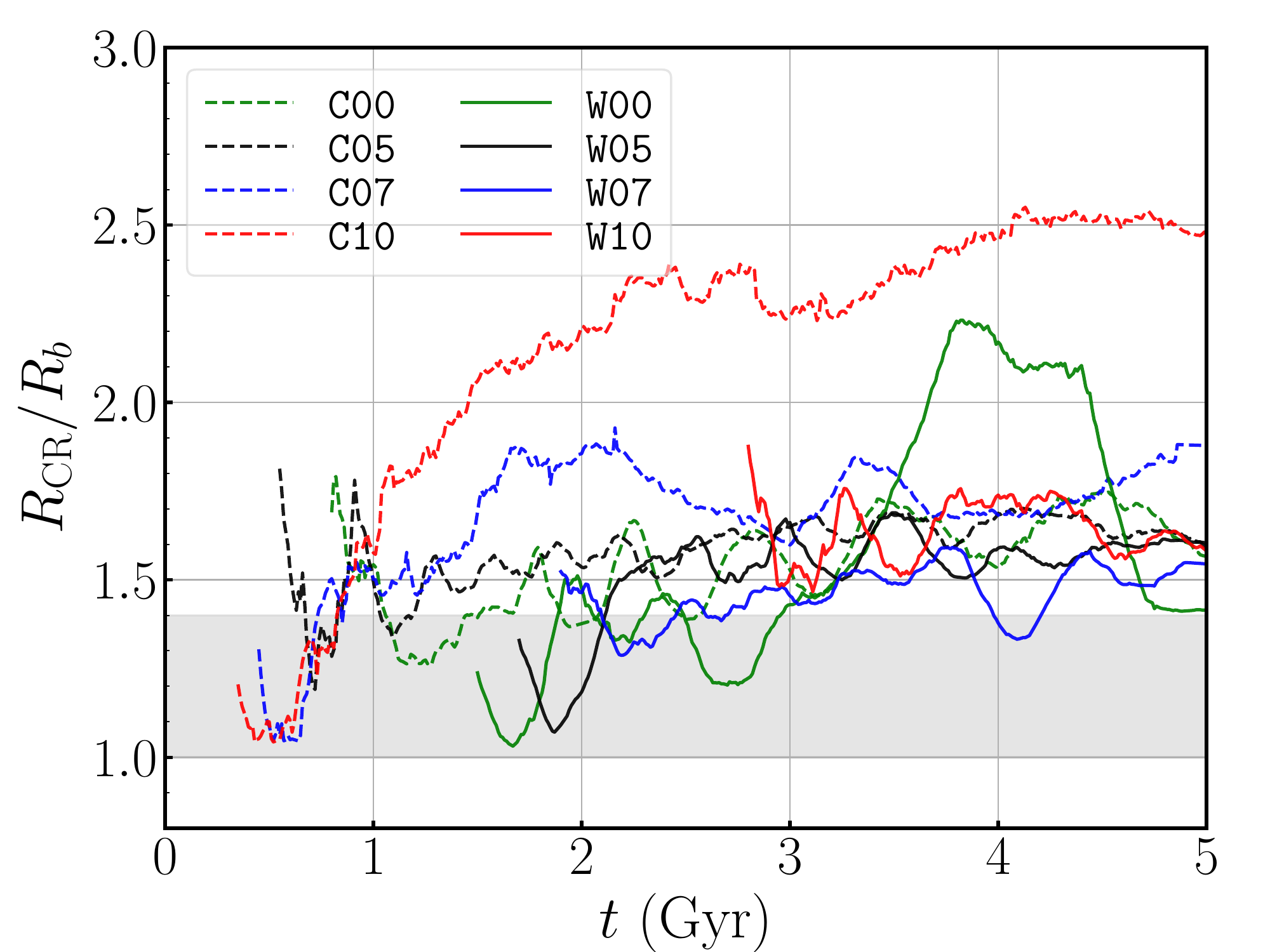}
\caption{
Temporal changes of $\mathcal{R}\equiv R_{\rm CR}/R_b$
for bars formed in the warm disks (solid) and cold disks (dashed).
The shaded regions correspond to
$1\leq \mathcal{R}\leq 1.4$. Except for a brief period in the formation stage, all bars in our models are slow rotators.
\label{fig:r_val}}
\end{figure}

Bars are classified as being ``fast'' or ``slow'' if $\mathcal{R}=R_{\rm CR}/R_b$ is less or greater than 1.4, respectively, where $R_{\rm CR}$ is the corotation radius.
Figure \ref{fig:r_val} plots temporal variations of
$\mathcal{R}$ for all the bars formed in our models.
The bars have $\mathcal{R}>1$ for all time, indicating that they
are located inside the corotation resonance.
The initial decrease of $\mathcal{R}$ is due to the rapid growth
of $R_b$ during the bar formation, resulting in fast bars
right after the formation.
The bars subsequently slow down by transferring
angular momentum to the surrounding halos, and turn
to slow rotators with $\mathcal{R} > 1.4$.
This is overall consistent with the previous results that bars in
simulations are usually slow unless disks
are highly gas-rich \citep{ath14} or the galaxy rotation curves
are dominated by a strong bulge \citep{pet18}.
Column (5) of Table \ref{tbl:model} gives the time-averaged values
of $\mathcal{R}$ over $t=4.5$--$5.0\Gyr$.
In model {\tt C10}, the bar evolves to an oval with $\mathcal{R}>2$.
The bar in model {\tt W00} undergoes buckling instability at $t\sim3.5\Gyr$ to become shorter (see below), resulting in $\mathcal{R}>2$ over
$t\sim3.5$--$4.5\Gyr$. Other than these,
the bars in our models have $\mathcal{R} \sim 1.5$--$1.8$
during most of their evolution.

\begin{figure}
\epsscale{1.0} \plotone{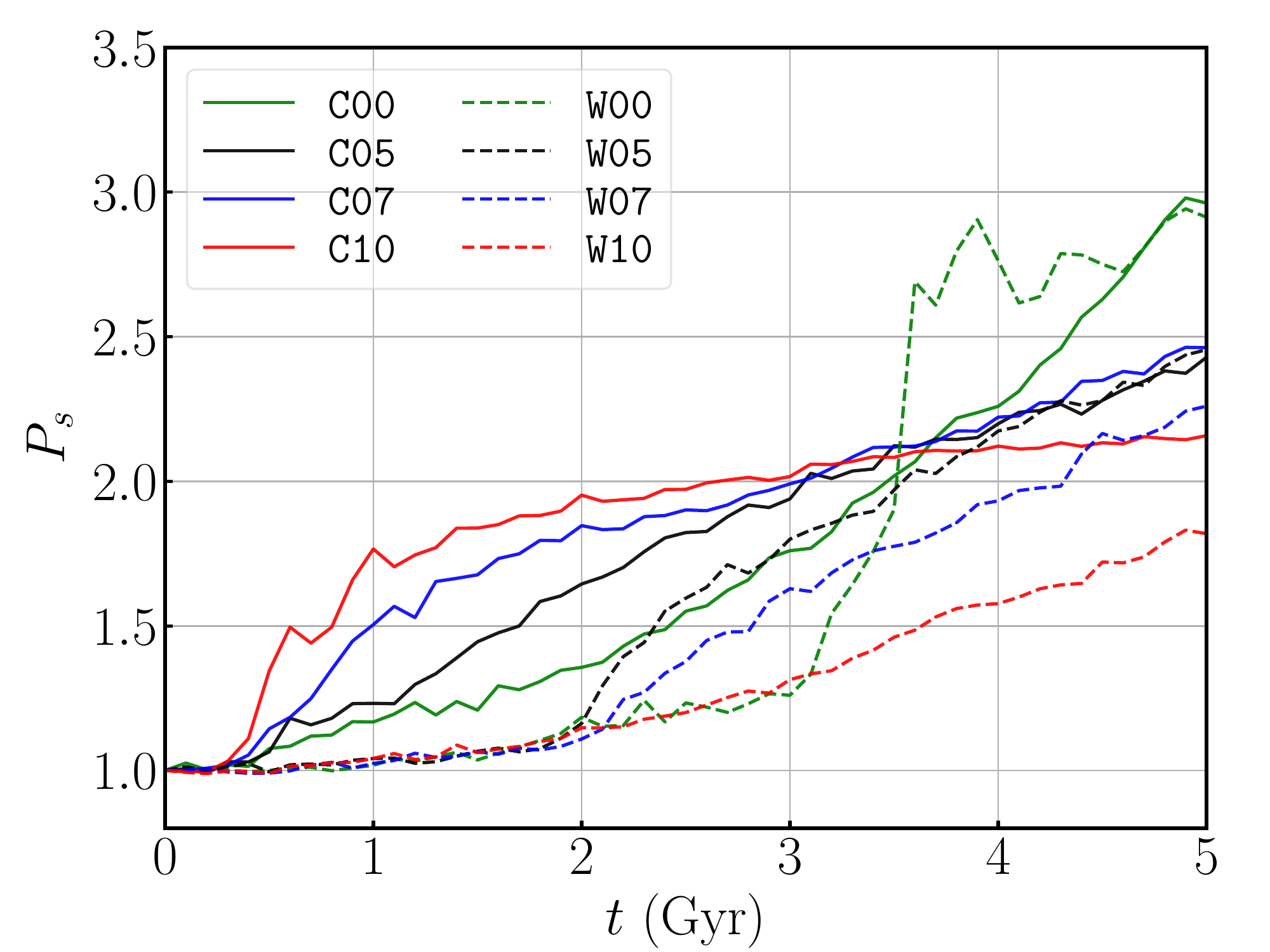}
\caption{
Temporal variations of the B/P strength, $P_s$, defined
as the maximum value of the median height of the stellar particles
relative to the initial value. In our models,
$P_s$ increases gradually due to the bar formation and enhancement
in the CMC and the bar mass, except model {\tt W00} that undergoes
a buckling instability to increase $P_s$ rapidly at $t\sim 3.5\Gyr$.
\label{fig:bps}}
\end{figure}

\begin{figure*}
\epsscale{1.2} \plotone{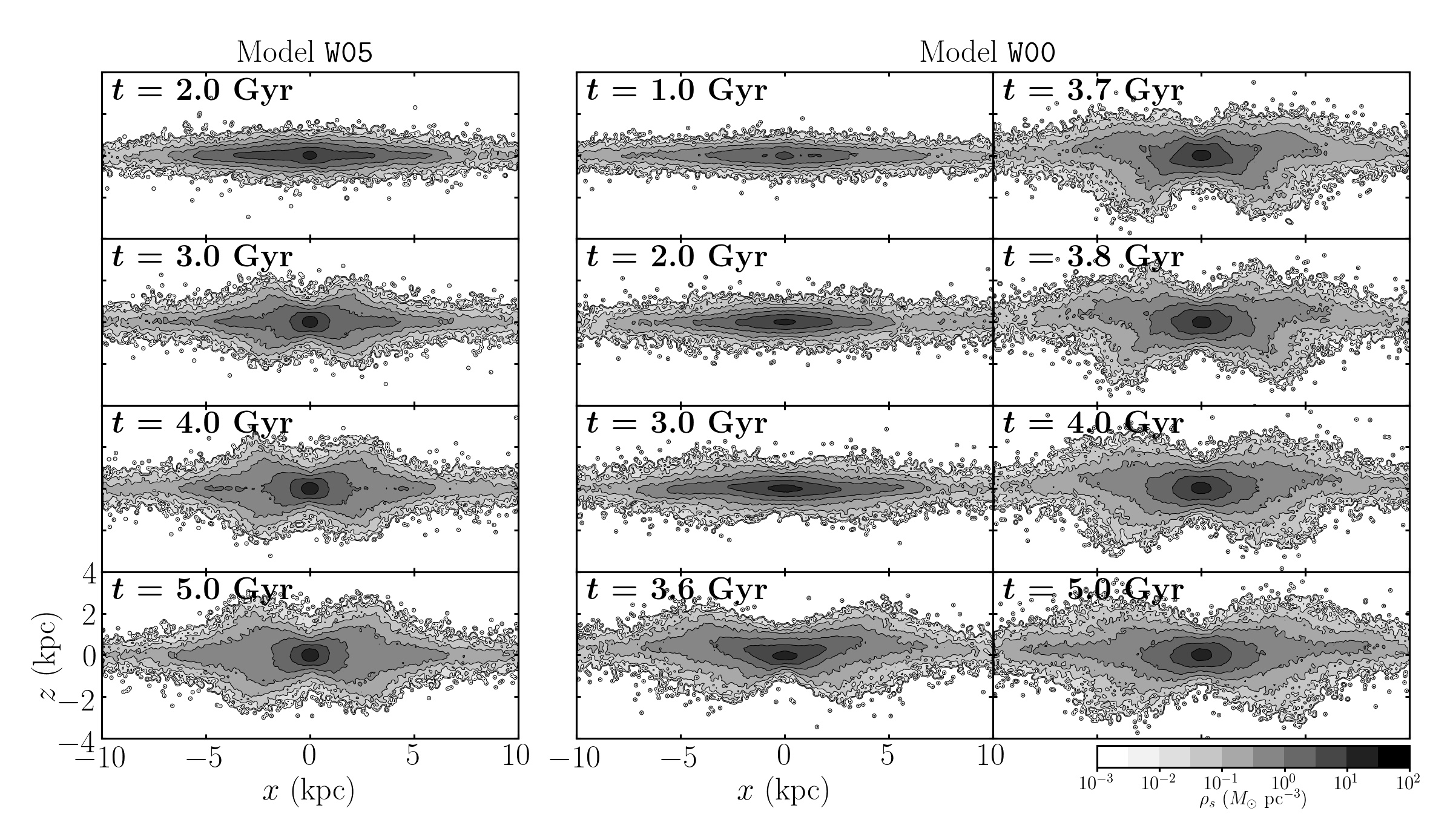}
\caption{
Contours of logarithm of the stellar density in the $x$-$z$ plane along the bar semi-major axis for models {\tt W05} and  {\tt W00} at selected times. The disk in model {\tt W05} thickens gradually to develop a B/P bulge, while the disk in model {\tt W00} undergoes buckling instability at $t=3.5\Gyr$ to become asymmetric with respect to the $z=0$ plane. 
\label{fig:buck}}
\end{figure*}

\begin{figure*}
\epsscale{1.0} \plotone{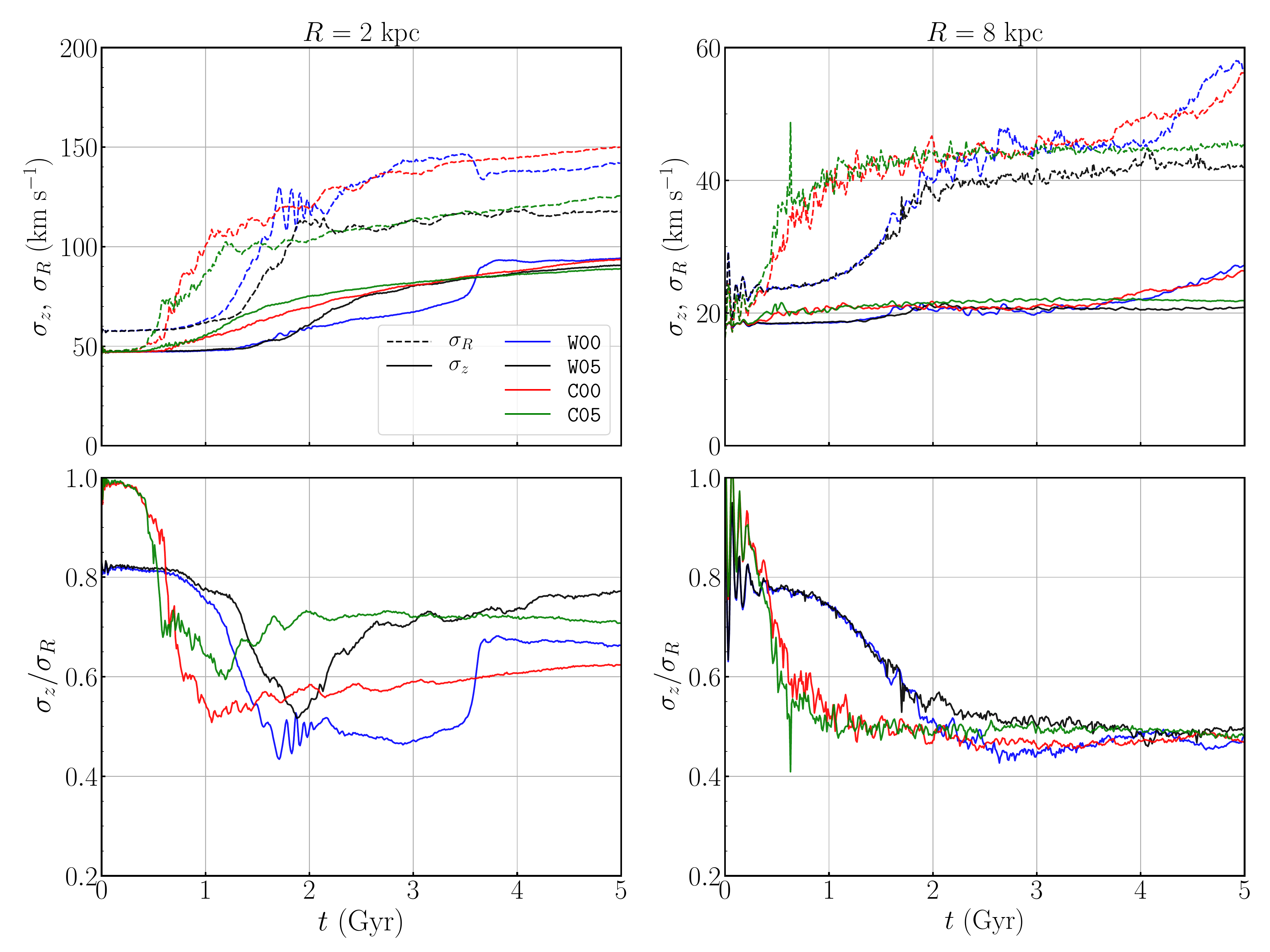}
\caption{
Temporal variations of the radial and vertical  velocity dispersions (top), and their ratio (bottom) for the models {\tt C00}, {\tt C05}, {\tt W00}, and {\tt W05} at $R=2\kpc$ (left) and $R=8\kpc$ (right).
\label{fig:disp}}
\end{figure*}

\subsection{Boxy/Peanut Bulge and Buckling Instability}\label{sec:buck}

We find that all bars in our models thicken over time and evolve to B/P bulges, except an oval in model {\tt C10} that remains in an ellipsoidal shape.
The thickness of a bar or boxy/peanut bulge is conventionally
measured by the B/P strength defined as
\begin{equation}\label{e:bps}
 P_s \equiv
  \max\left\{\frac{\widetilde{|z|}}{ \widetilde{|z_0|}}\right\},
\end{equation}
where the tilde indicates the median and the subscript ``0" refers to the initial disk configuration (e.g., \citealt{ian15,fra17}).
Figure \ref{fig:bps} plots the temporal variations of the B/P strength
in our models. The B/P strength increases gradually over time due
to the CMC as well as the bar mass, both of which heat the disk by exciting stellar motions along the vertical direction.  The late-time increase of $P_s$ in model {\tt C00} is largely caused by the increase in the bar mass rather than the CMC.
The maximum thickness occurs typically at $R\sim 2$--$3\kpc$, with larger values corresponding to stronger bars.
In addition to this gradual thickening, model {\tt W00} also experiences a rapid increase of $P_s$ at $t\sim 3.5 \Gyr$, which is caused by a vertical buckling instability.

To illustrate the buckling instability,
Figure \ref{fig:buck} plots the contours of the stellar density
in the $x$--$z$ plane at some selected epochs of
models {\tt W00} and {\tt W05},
where $x$ and $z$ denote the directions along
the bar semi-major axis and perpendicular to the galactic plane, respectively.
The disk (or bar) in model {\tt W05} thickens gradually
along the vertical direction and remains
almost symmetric with respect to the midplane ($z=0$) during its
entire evolution. On the other hand, the disk in model {\tt W00}
thickens gradually for $t\lesssim3\Gyr$ and
then undergoes a fast buckling instability at $t=3.5\Gyr$,
promptly increasing $\sigma_z$ and breaking the reflection symmetry
about the $z=0$ plane (e.g., \citealt{mar06}).   It also causes a temporal drop in the bar strength (see Figure \ref{fig:str_cmc}). The disk subsequently becomes more-or-less symmetric at $t=4\Gyr$, and the bar becomes a B/P bulge.

The operation of buckling instability requires the ratio
$\sigma_z/\sigma_R$ of the velocity dispersions to be smaller than
a critical value (e.g., \citealt{bin07}).
\citet{too66} and \citet{ara87} found that infinitesimally-thin, non-rotating slabs are unstable if $\sigma_z/\sigma_R < 0.3$, while \citet{mer94} suggested an instability criterion of $\sigma_z/\sigma_R<0.6$ for axisymmetric
rotating disks. For barred disks with spatially varying $\sigma_z/\sigma_R$, \citet{mar06} and \citet{kwa17} used $N$-body simulations to show that the critical values for buckling are at
$\sigma_z/\sigma_R\sim0.6$ for their models,
suggesting that the critical value may depend
on the density and velocity distributions inside the disk.

Figure \ref{fig:disp} plots temporal changes of
$\sigma_z/\sigma_R$ for models {\tt C00}, {\tt C05}, {\tt W00}, and {\tt W05} at $R=2\kpc$ (left) and $R=8\kpc$ (right).
The bar formation in itself increases $\sigma_R$, while
the CMC and the bar mass tend to increase $\sigma_z$. Since the bar formation primarily involves the mass re-distribution in the galactic plane, $\sigma_R$ increases more rapidly and strongly than $\sigma_z$.
In the inner disk where a bar is located,
this causes $\sigma_z/\sigma_R$ to decrease with time in the early phase of bar evolution ($t<1$ and $2\Gyr$ for the cold- and warm-disk models), and subsequently to increase as the increase of $\sigma_R$ slows down.

The minimum value of $\sigma_z/\sigma_R$ is determined by the competition between the bar strength and the CMC.
It turns out that all of our models with gas included suffer a large
increase in $\sigma_z$ to always have $\sigma_z/\sigma_R > 0.5$
in the bar regions throughout their entire evolution,
and thus remain stable to the buckling instability.
Although model {\tt C00} has no gas, its bar is strong enough to heat the disk vertically, resulting in $\sigma_z/\sigma_R > 0.5$ for all time. On the other hand,
the bar in model {\tt W00} grows strong but relatively slowly, and
incurs only a mild increase in the CMC and the bar mass. As a consequence, it has $\sigma_z/\sigma_R$ as low as $\sim0.47$
and undergoes buckling instability. These results suggest that the critical value for the buckling instability is $\sigma_z/\sigma_R\sim0.5$
for our models, and that the presence of gas tends to suppress
buckling instability by increasing the CMC.
This is qualitatively similar to the results of \citet{ian15} who found that models without gas belong to a strong-B/P group where
bars undergo bucklings and result in high $P_s$, while
those with a large fraction of gas involve gradual bar and $P_s$ growth and are in a moderate-B/P group.
\citet{ber07} also showed that gas-free/poor disks experience
buckling,  while gas-rich models thicken due to vertical heating
instead of buckling.

The right panels of Figure \ref{fig:disp} show that the bar formation
increases $\sigma_R$ significantly also in the outer regions. However, the excitation of the vertical stellar motions due to the CMC and the bar mass is negligible in these regions, making $\sigma_z$ almost unchanged. The amount of the increment in the velocity dispersions is insensitive to the gas fraction until $t\sim 4 \Gyr$ when the bar grows in size sufficiently to directly influence stellar orbits in the outer regions. This makes
$\sigma_z/\sigma_R$ at $R=8\kpc$ reduced to $\sim0.45-0.5$ at the end of the runs, almost independent of $\fgas$. The velocity dispersions at $t\gtrsim 2\Gyr$ in both cold- and warm-disk models are consistent with
the solar neighborhood values $\sigma_R\sim 40$--$50\kms$ and $\sigma_z\sim 25$--$35\kms$ of the Milky Way obtained from the analysis of the second \emph{Gaia} data release (e.g., \citealt{kat18}; see also \citealt{sha14,gui15}).

\begin{figure*}
\epsscale{1} \plotone{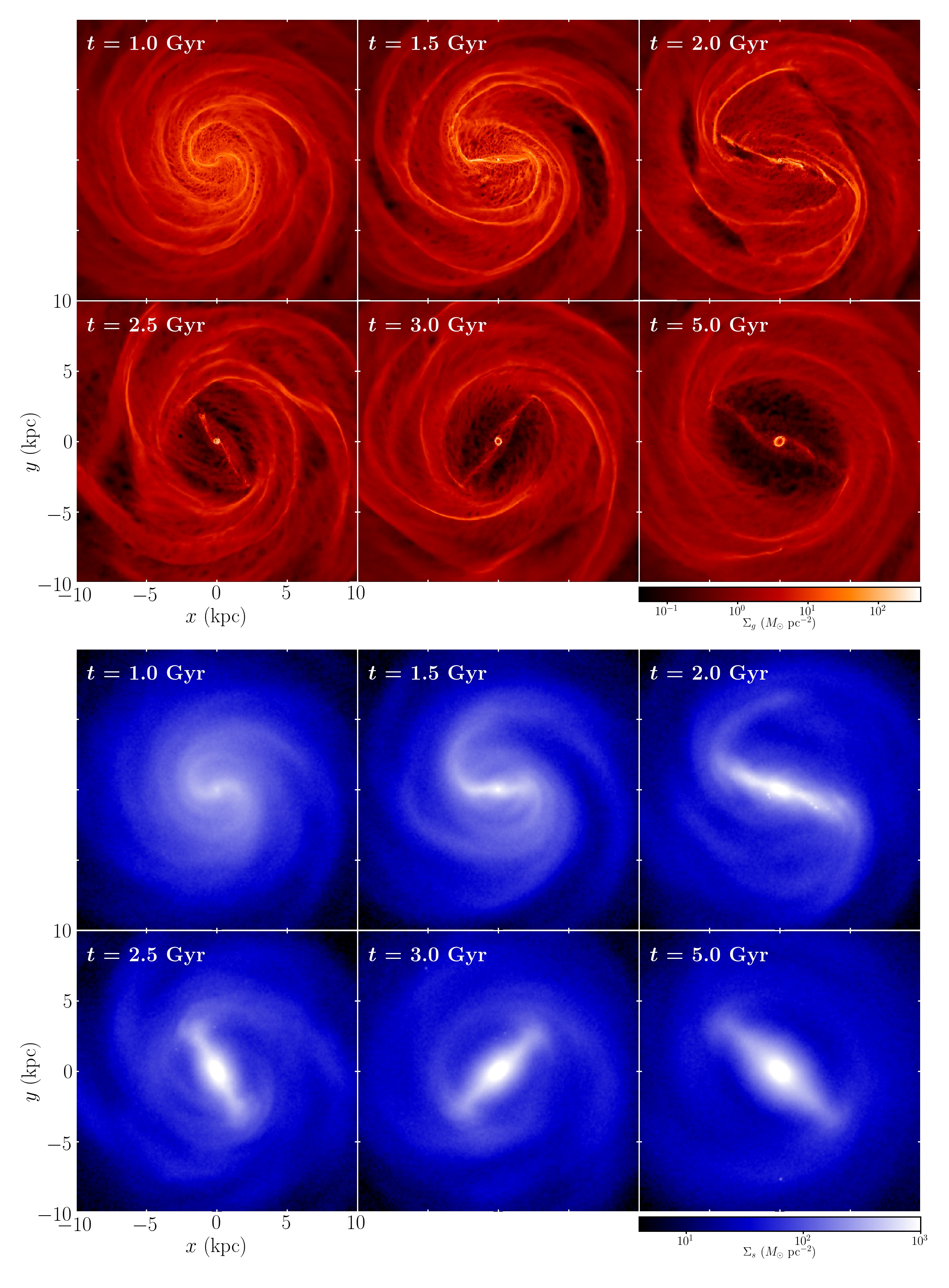}
\caption{
Snapshots of logarithm of the surface density of the gaseous component (upper) and the stellar component (lower) for model {\tt W05} at six epochs in the $10\kpc$ regions.
In the lower panels, the stars include both the pre-existing
particles in the initial stellar disk and the new stellar particles
converted from the gaseous disk.
\label{fig:snap_g10}}
\end{figure*}

\begin{figure*}
\epsscale{1} \plotone{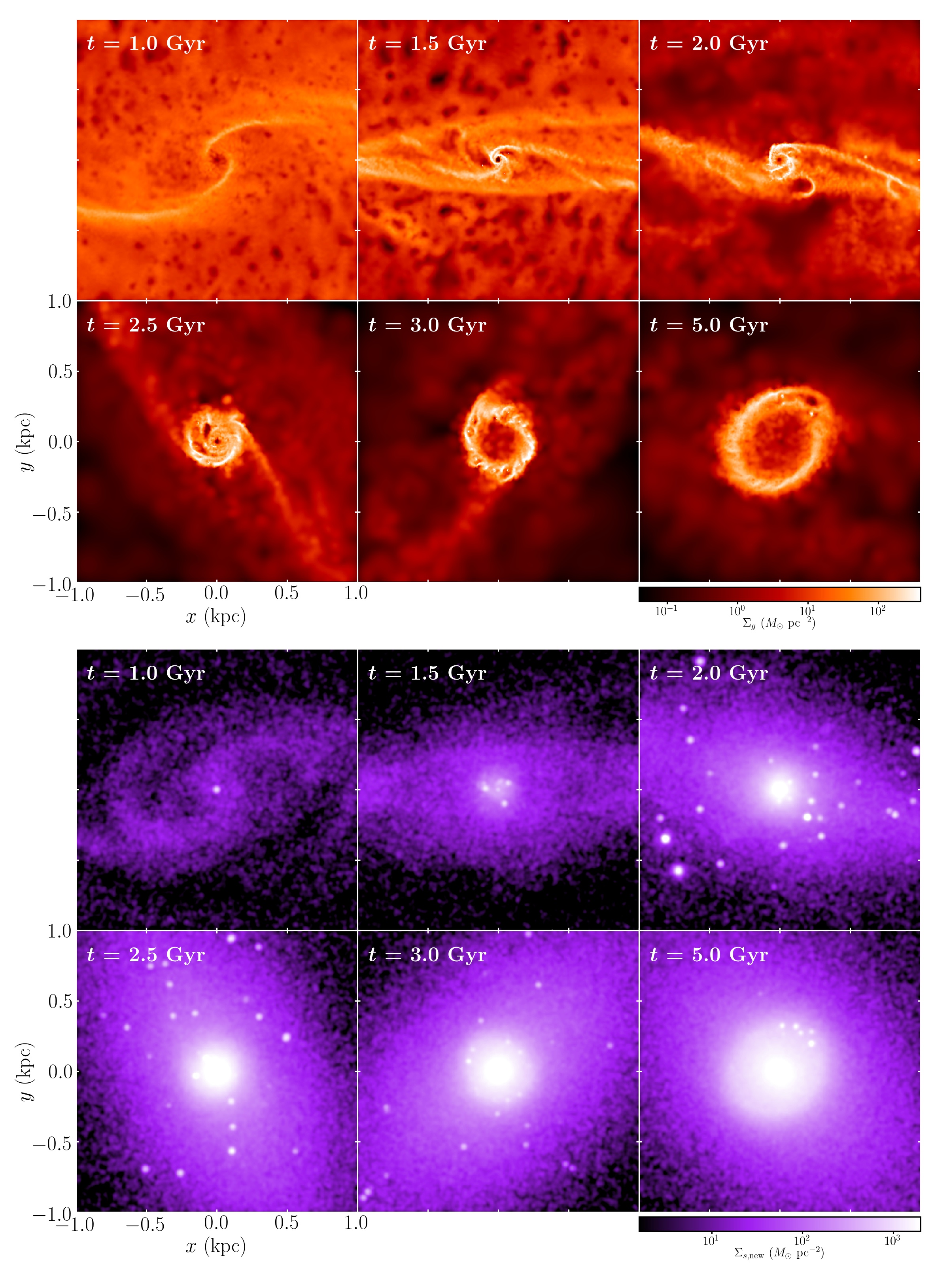}
\caption{
Same as Figure \ref{fig:snap_g10}, but for the central $1\kpc$ regions. In the lower panels, the stars represent only
the new stellar particles converted from the gaseous disk.
\label{fig:snap_gin}}
\end{figure*}

\section{Gaseous Structures and Star Formation}\label{sec:gas}

\subsection{Spiral Structures}\label{sec:spiral}

Figure  \ref{fig:snap_g10} plots snapshots of the gas surface density (upper panels) as well as the stellar surface density (lower panels)
in logarithmic scale of model {\tt W05} in the $10\kpc$ regions at six different epochs. The stellar surface density counts both the pre-existing particles in the initial stellar disk and the new stellar particles created from the gaseous disk.
Both the bars and gas are rotating in the counterclockwise direction.

In the regions outside the bar, the gas disk exhibits spiral structures that not only trace the local minima of the total gravitational potential but also are sites of star formation.
The location of the gaseous spiral arms almost coincides with that of the stellar spiral arms, although the former is narrower and more strongly peaked. The number of arms depends on $f_R$ and $\fgas$ when they form, and their pattern speed closely follows the local angular speed $\Omega$ of galaxy rotation (see Figure \ref{fig:ncc}). This suggests that the arms have a character of material arms at early time, similarly to self-generated arms driven by swing amplifications
(e.g., \citealt{bab13,don13}). They then merge themselves and start to interact with the bar potential that is growing, eventually ending up being four spirals that are piecewise logarithmic in shape, with a pitch angle of $\sim9^\circ$--$12^\circ$.  Figure \ref{fig:ncc} shows that at late time the spirals tend to roughly corotate with the bar, which is consistent with the results of \citet{roc13} who showed that spiral arms in unbarred disks have a pattern speed close to $\Omega$, whereas those in barred-spiral models corotate with a bar.

\subsection{Nuclear Ring}\label{sec:ring}

\begin{figure*}
\epsscale{1.0} \plotone{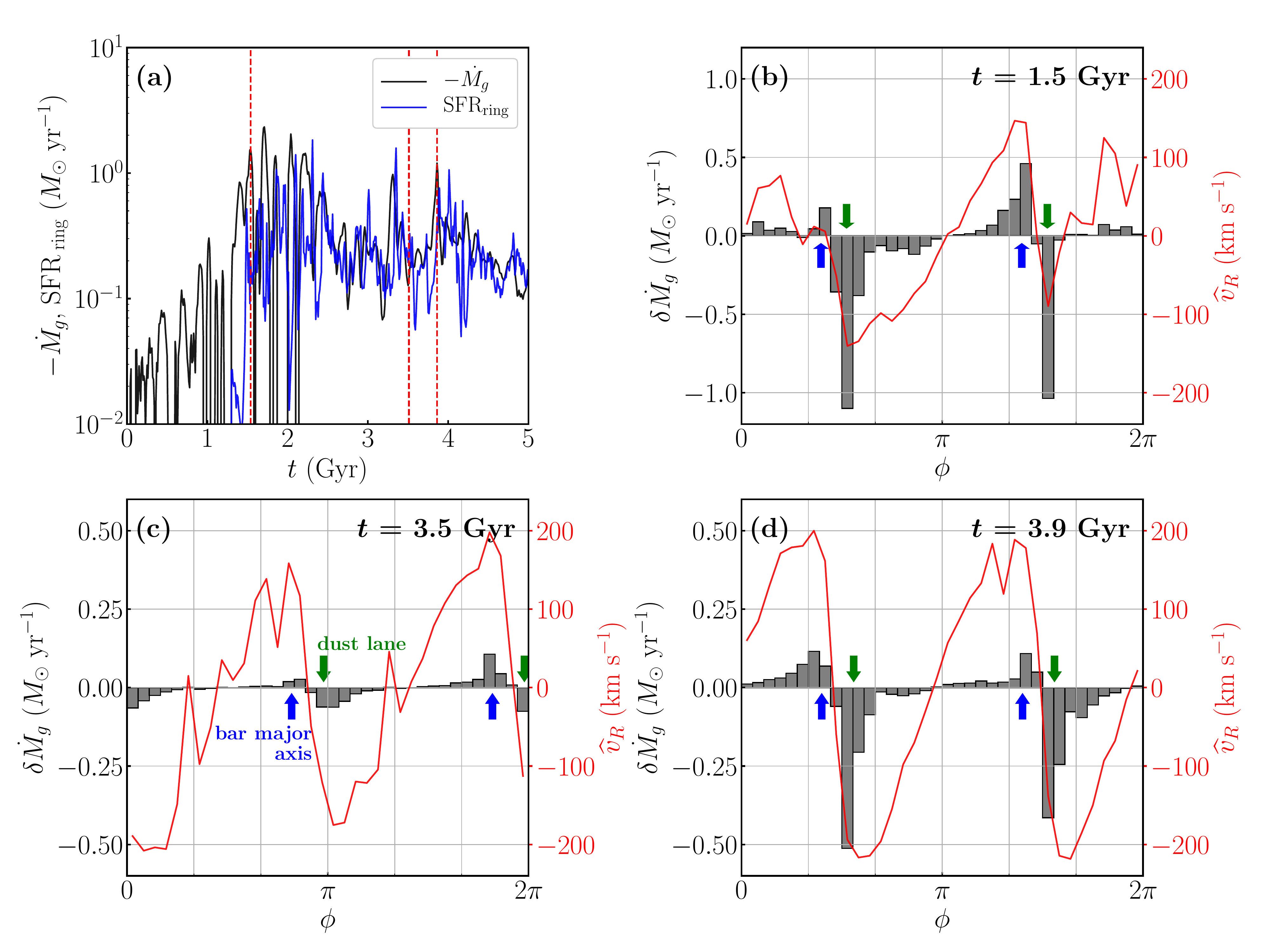}
\caption{
(a) Temporal variations of the gas inflow rate $-\dot{M}_g$ measured at $R=1\kpc$ (black), compared with the ring star formation rate SFR$_{\rm ring}$ (blue) for model {\tt W05}.
(b)--(d) Angular distributions of the gas inflow rates
$\delta \dot{M}_g$ in an azimuthal bin with size
$\delta \phi=10^\circ$ (histograms;
left axis) and the density-weighted radial gas velocity
$\widehat{v}_R (R=0.5\kpc)$ (red line; right axis)
at $t=1.5$, $3.5$, and $3.9\Gyr$ of model {\tt W05}, which are marked as dashed vertical lines in (a).
The arrows in blue and green indicate the locations
of the bar semi-major axis and the dust lanes,
respectively.
\label{fig:minfall}}
\end{figure*}

\begin{figure}
\epsscale{1} \plotone{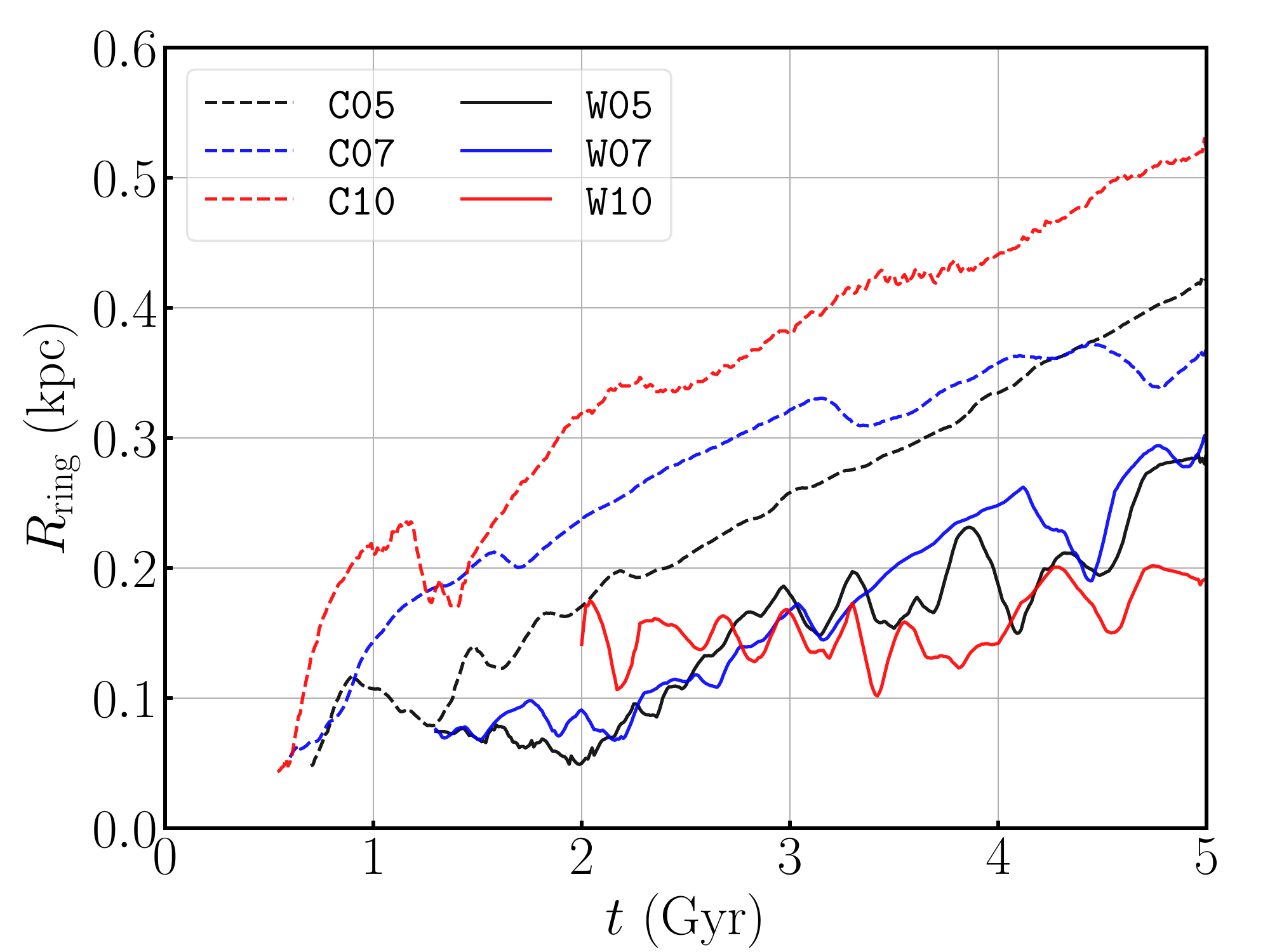}
\caption{
Temporal variations of the ring size for all models with gas.
Nuclear rings are small when they form, and grow in size over time as
the bars become longer.
\label{fig:ring_size}}
\end{figure}

The regions inside $R=R_b$ are governed by the bar potential.
In model {\tt W05}, a weak stellar bar at $t=1\Gyr$ produces a pair of dust lanes that are relatively straight at $R\gtrsim 0.5\kpc$ and
take a form of trailing spirals toward the center (e.g., \citealt{mac04}).
As the bar grows further, the dust lanes become quite straight and are located very close to the semi-major axis of the bar (e.g., \citealt{kim12a}).
To display detailed structures of dust lanes and a nuclear ring,
Figure \ref{fig:snap_gin} zooms in the central $1\kpc$ regions of
the snapshots shown in Figure \ref{fig:snap_g10}, with the lower
panels showing only the stars produced from the gaseous disk.
At $t=1.0\Gyr$, there are some stars formed from the dust lanes
that are curved, yet no ring is created at the center.
A small nuclear ring with radius $\sim40\pc$ is
beginning to form at $t=1.5\Gyr$ as the material driven
inward by the bar potential accumulates near the galaxy center.

To quantify the mass inflow, we calculate the binned
gas inflow rate $\delta \dot{M}_g (R,\phi_i)=
\int_{\phi_i}^{\phi_i+\delta \phi}
\Sigma_g v_R R d\phi$,
where $\phi_i$ and $\delta \phi=10^\circ$ denote the bin boundaries and width, respectively. The gas inflow rate at $R$ is then
given by $\dot{M}_g (R)= \sum_i \delta \dot{M}_g(R,\phi_i)$.
Figure \ref{fig:minfall}(a) plots as a black solid line the temporal changes of $-\dot{M}_g$ at $R=1\kpc$ in model {\tt W05}. For comparison, we also plot the ring star formation rate SFR$_{\rm ring}$ as a red line, which will be discussed in Section \ref{sec:sfr}. Figure \ref{fig:minfall}(b-d) plots the angular distributions of the binned mass inflow rate $\delta \dot{M}_g$ (histograms) as well as the vertically averaged radial velocity
$\widehat{v}_R=\int \rho_g v_R dz /\Sigma_g$ (solid lines)
at $R=0.5\kpc$ and $t = 1.5$, $3.5$, and $3.9 \Gyr$
for model {\tt W05}. These times are chosen to illustrate the cases when the bar is growing ($t=1.5\Gyr$), when the mass inflow rate is small ($t=3.5\Gyr$)
or large ($t=3.9\Gyr$) after the bar reaches roughly a quasi-steady state.
Clearly, the gas infall occurs mainly along the dust lanes (green arrows) located downstream from the bar semi-major axis (blue arrows).
While $\dot{M}_g$ fluctuates with large amplitudes during the
bar formation due to star formation feedback,
it varies quite mildly around the mean value of about $-0.25\Aunit$
after $t\sim2\Gyr$, and the maximum binned inflow rates
amount to $\delta \dot{M}_g = -0.1$ to $ -0.5 \Aunit$
with inflow velocities of $\widehat{v}_R = 150$--$200\kms$.
Since the gas in the bar regions more-or-less follows
$x_1$-orbits, some gas in the upstream side from the semi-major axis moves outward radially, but the associated outflow rate is in general smaller than the inflow rate, resulting in overall inflows.

To explore how the ring size varies with time in our models, we measure the density-weighted, angle-averaged ring radius using
\begin{equation}\label{e:ring}
R_{\rm ring} \equiv
  \left. \frac{\int{R\Sigma_g(R,\phi) dR d\phi}}{\int{\Sigma_g(R,\phi) dR d\phi}}
  \right|_{\rm max},
  \quad\text{for}\;R\leq 1\kpc.
\end{equation}
Figure \ref{fig:ring_size} plots $R_{\rm ring}$ as functions of time for all models with gas.
In model {\tt W05}, the ring remains small and exhibits small fluctuations until the bar strength reaches its peak at $t=2\Gyr$.
During this time, intermittent star formation occurring
in the nuclear regions disperse the ring and dust lanes temporarily.

\begin{figure*}
\epsscale{1} \plotone{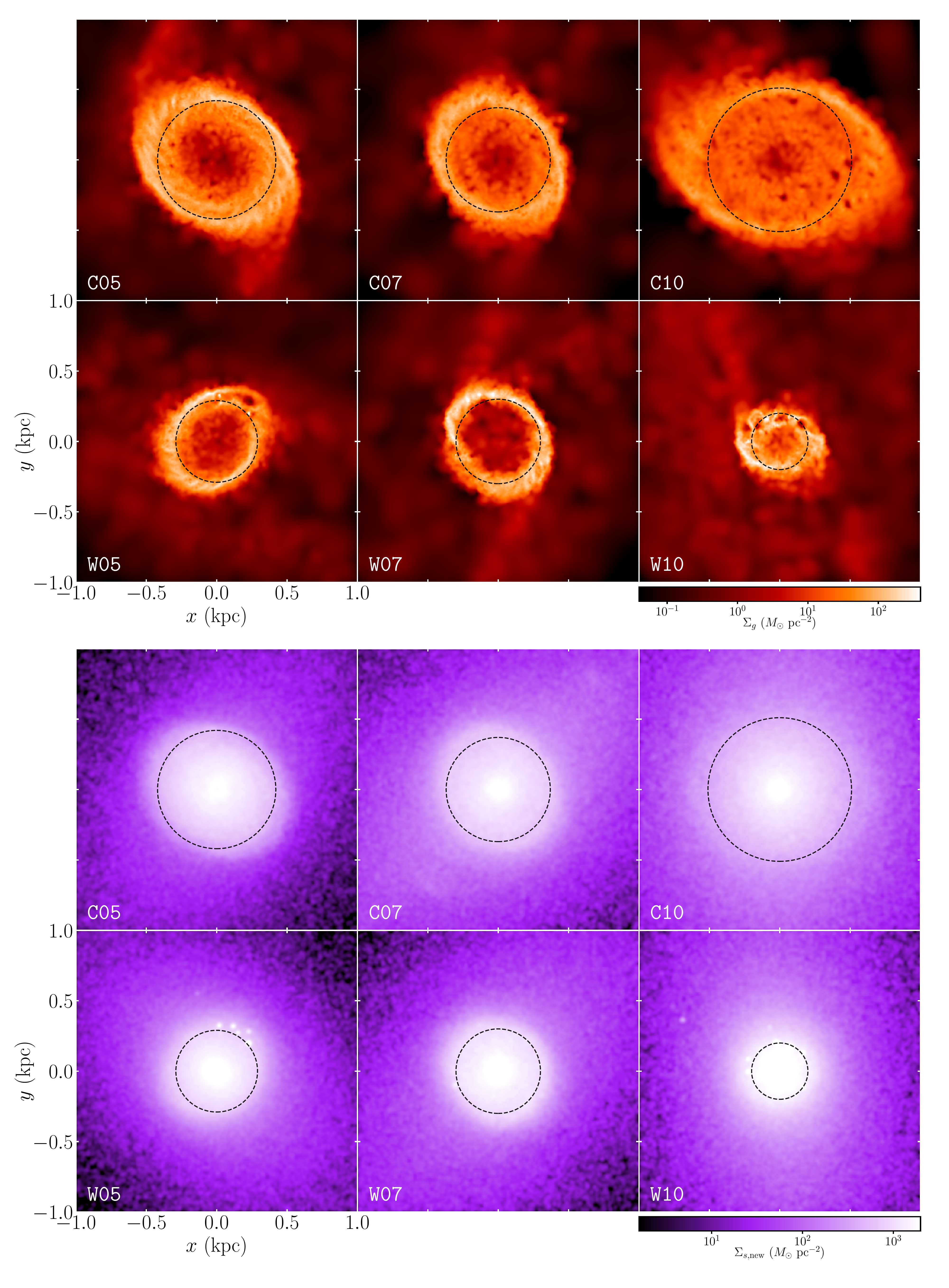}
\caption{
Snapshots of logarithm of the gas surface density (upper panels)
and the surface density of the newly formed stars (lower panels)
at $t=5\Gyr$ in the $1\kpc$ regions for all models with gas.
The dashed circle in each panel draws the ring size calculated from
Equation \eqref{e:ring}.
\label{fig:ring_5Gyr}}
\end{figure*}

After the bar achieves the maximum strength, the gas in the bar regions experiences massive infall and is added
continuously to the ring.  The ring
in model {\tt W05} grows in size slowly with time to $R_{\rm ring}\sim 0.3\kpc$ at $t=5\Gyr$, which is caused primarily by the
increase in the bar size (Fig.~\ref{fig:length}a). As the bar becomes longer, fresh gas at larger $R$ and thus with higher angular momentum infalls to be added to the ring. At the same time, the gas already in the ring that has lower angular momentum than the gas being added is continually consumed to
star formation at a rate of $\sim0.2\Aunit$.
Since the ring mass is typically $\sim 4\times 10^7\Msun$, the ring gas is almost completely replaced by newly inflowing gas from outside
in $\sim0.2\Gyr$.
The decrease in the bar pattern speed which tends to make the dust lanes move away from the bar semi-major axis (e.g., \citealt{li15}) as well as the CMC increase also help to increase the ring size.
Large fluctuations in the ring size at late time in the warm-disk models are due to active star formation feedback (see Section \ref{sec:sfr}).
Column (6) of Table \ref{tbl:model} lists the time-averaged values of
$R_{\rm ring}$ over $t=4.5$--$5.0\Gyr$.
Although all rings in our models keep growing until the end of the simulations,
they would stop growing when the bars reach a steady state.


Figure \ref{fig:ring_5Gyr} compares the distributions of the gas surface density (upper panels) and the surface density of the newly formed stars (lower panels) at $t=5\Gyr$ in the nuclear regions in all models with gas. The black dashed circles mark the ring size obtained by Equation \eqref{e:ring}.
Although the newly formed stars as a whole are smoothly distributed at $t=5\Gyr$ without noticeable features, we find that young stars with age less than $\sim 0.5\Gyr$ are in a ring shape, similarly to the gaseous counterpart, which may be observed as stellar nuclear rings in the TIMER survey (e.g., \citealt{gad19}).
Stars older than this age were also in a ring at earlier time, but
their radial diffusion through mutual scattering makes it difficult
to keep them in a ring shape.

Overall, the rings in
the cold disks are larger than in the warm disks since the former forms earlier and thus can grow for a longer period of time until the end of the runs. There is no apparent correlation between the ring size and the gas fraction. In model {\tt W10}, the nuclear ring has a radius of $R_{\rm ring} \sim0.15\kpc$,
largest among the rings, when
it first forms at $t\sim2\Gyr$, and its size does not vary much with time afterward. This is because the CMC that is already sufficiently strong at the time of ring formation
overwhelms the effect of the bar growth that occurs quite slowly
(e.g., \citealt{li17}).
Although the bar (or oval) in model {\tt C10} remains short and weak throughout most of its evolution, the gas inflows driven by spiral arms into the bar regions are
significant to make the ring larger in size over time (e.g., \citealt{kk14,seo14}).

That a nuclear ring forms small and becomes larger with time
found in our current models is
different from the cases with a static bar potential
in which a ring is large when it forms (e.g., \citealt{kim12a,kim12b}).
When the properties of a stellar bar are fixed with time, the non-axisymmetric bar torque produces dust-lane shocks downstream away from the bar semi-major axis. In this case, a ring forming at the inner ends of the dust lanes has quite a large radius, although it subsequently shrinks in size
by 10 to 20\% as collisions of dense clumps inside the ring take away angular momentum from the ring \citep{kim12a}.
In our self-consistent models, on the other hand, the physical properties of stellar bars
keep changing with time, providing non-steady gravitational potentials to the gas.
Near the time when a nuclear ring forms ($t\sim1.5\Gyr$),
the bar potential is strong enough to induce shocks
only in the innermost ($R\lesssim0.1\kpc$) parts of the dust lanes,
so that the resulting nuclear ring should be much smaller than
the counterpart under the fixed bar potential where the dust-lane
shocks are extended across the whole bar length.

\begin{figure*}
\epsscale{1.1}
\plotone{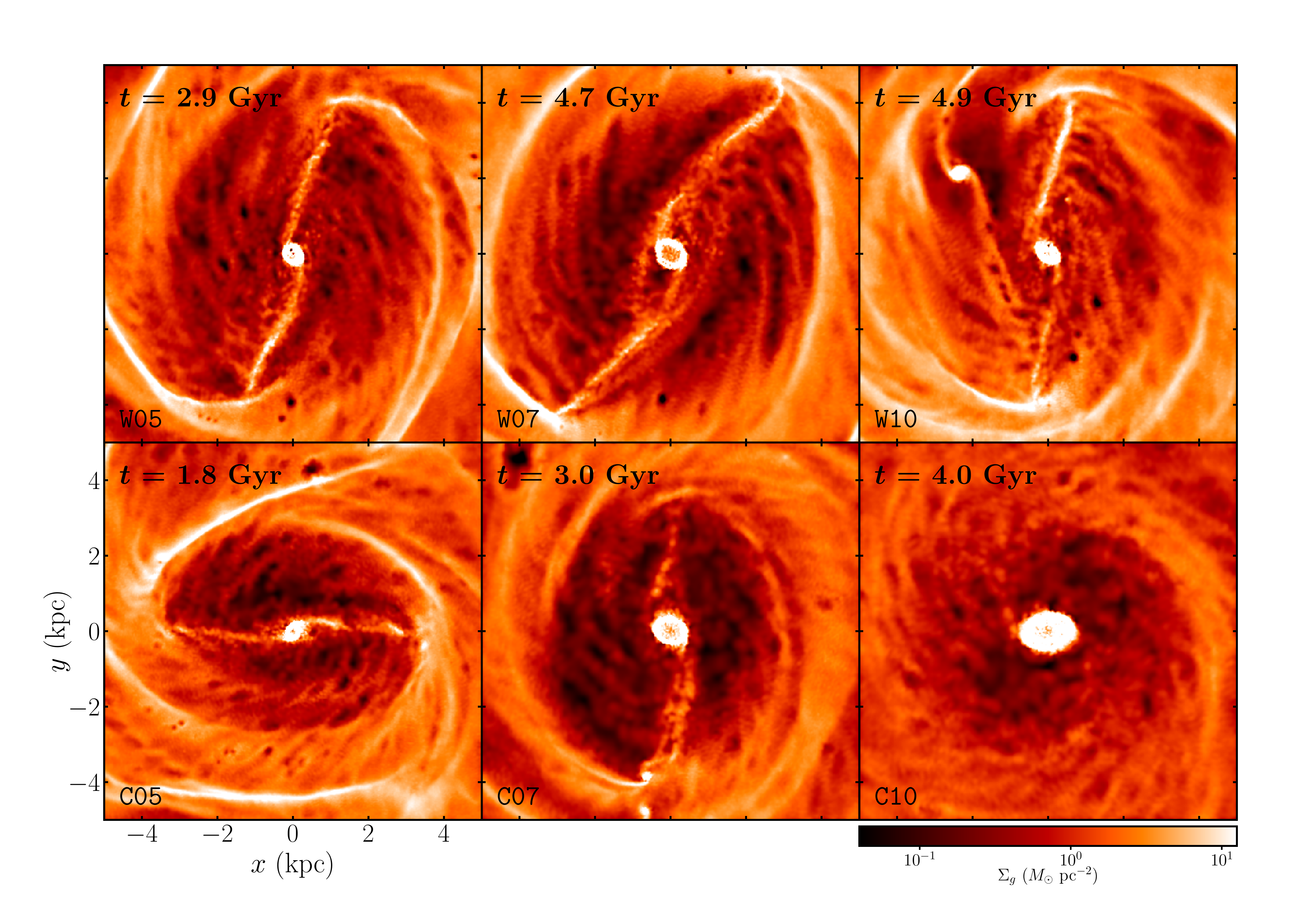}
\caption{
Logarithm of the gas surface density in the central $5\kpc$ regions for all models with gas. The selected time for each model is marked at the top left corner of each panel, while the model name is given at the lower left corner. Except in model {\tt C10} with an oval, interbar spur-like structures linked almost perpendicularly to the dust lanes are common.
\label{fig:spur}}
\end{figure*}

\subsection{Filamentary Spurs}\label{sec:spur}

While perpendicular filamentary interbar spurs have often been observed in association with dust lanes in real galaxies
(e.g., \citealt{she02,zur08,elm09}),
previous hydrodynamic simulations with a fixed bar potential
were unable to produce such structures
(e.g., \citealt{kim12a,kim12b,seo13,seo14}).
In these simulations, the bar regions quickly
reach a quasi-steady state in which gas approximately
follows $x_1$-orbits that are almost parallel to the dust
lanes (see, e.g., Figure 7 of \citealt{kim12b}).

Unlike in the previous simulations, we find that the current
self-consistent simulations with gas produce
interbar spur-like structures during the time when a bar grows,
except for model {\tt C10} in which a bar rapidly evolves to a weak oval without dust lanes.
Figure \ref{fig:spur} plots logarithm of the gas surface density in the inner $5\kpc$ regions
for all models with gas at selected epochs when spur-like structures are vivid. It is apparent that in all models except model {\tt C10}
several interbar spurs are connected perpendicularly
to the dust lane. These structures are transient, lasting typically for $\sim 0.1\Gyr$
before being destroyed by nearby star formation.
In terms of the number, shape, and density of spurs,
there is no noticeable difference from model to model.
The density enhancement associated with the spurs
is only $\sim1\Surf$ in the interbar regions, which becomes larger in
the dust lanes by up to an order of magnitude.
High-density peaks formed by collisions of the spurs with the dust lanes sometimes undergo star formation.

\begin{figure*}
\epsscale{1.0} \plotone{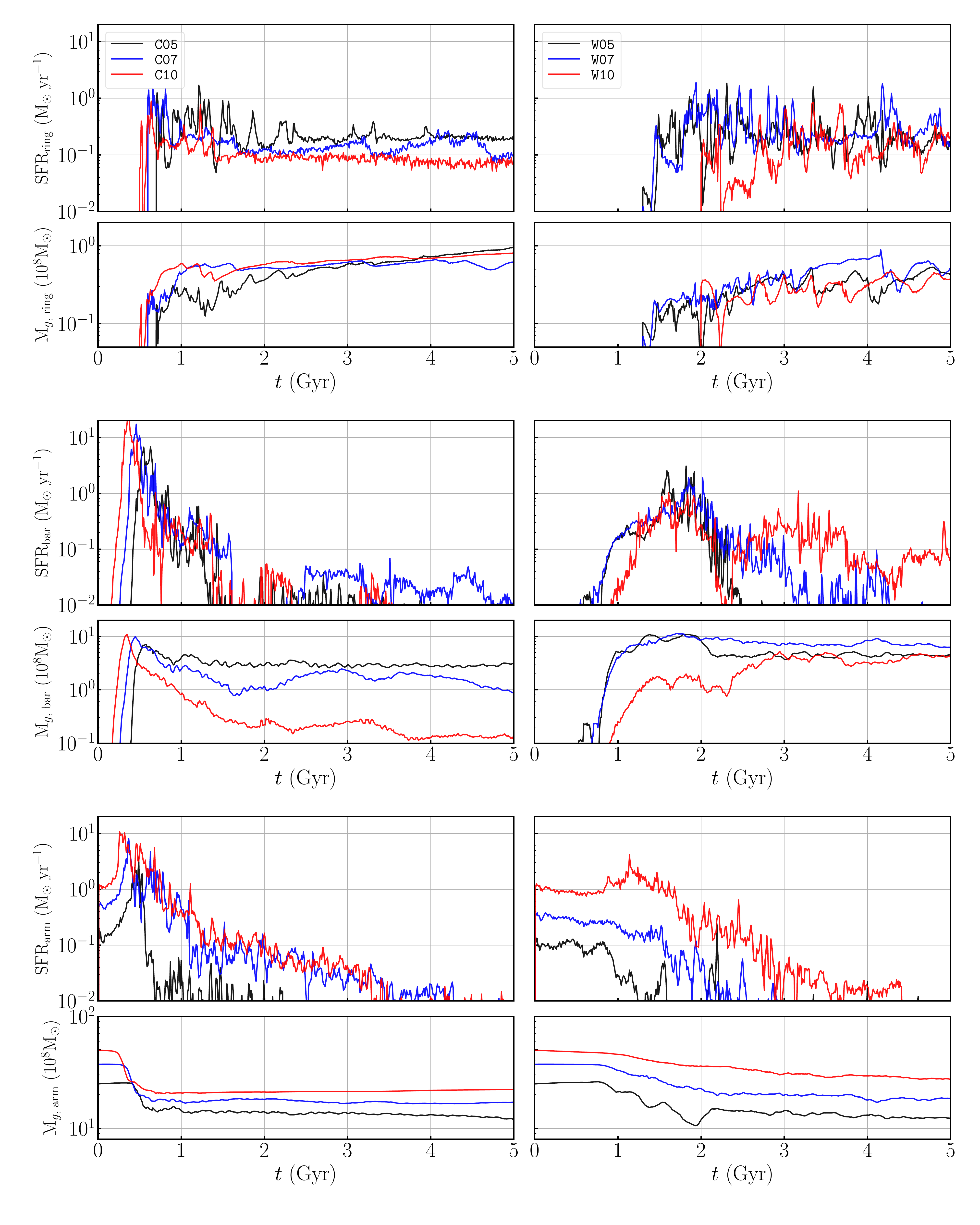}
\caption{
Temporal variations of the SFR and the gas mass in the
ring regions at $R<R_{\rm ring} + 0.2\kpc$ (top), the bar regions
at $R_{\rm ring} + 0.2\kpc<R<R_b + 0.5\kpc$ (middle), and the
arm regions at $R>R_b + 0.5\kpc$ (bottom). The left and right panels are for the cold- and warm-disk models, respectively.
\label{fig:sfr}}
\end{figure*}

Spur-like structures in our models are originated from star formation
feedback as well as non-steady gas streamlines.  Shells produced by
feedback in the low-density interbar regions are stretched by shear in the background flows, creating filamentary structures there.
Since the bars in our simulations change with time,
there are no well-defined $x_1$-orbit families
that gas can follow.  As the bar slows down, the gas velocity relative to the bar becomes larger. With the increased ram pressure,
the dust lanes slowly move away from the bar
semi-major axis  (e.g., \citealt{li15}),
and the gas streamlines that
turn their directions near the semi-major axis become almost perpendicular to the dust lanes.
The sheared filaments also turn directions near the
bar semi-major axis and hit the dust lanes perpendicularly to
enhance the local density.

\subsection{Ring Star Formation}\label{sec:sfr}

Star formation in our cold-disk models is widely distributed across the entire disk, while the warm-disk models actively form stars only inside the bar and nuclear regions. Using the instantaneous bar and ring sizes, we divide the entire disk into three parts: the arms regions with $R> R_b + 0.5 \kpc$, the bar regions with $R_b+0.5\kpc>R>R_{\rm ring}+0.2\kpc$, and the ring regions with $R< R_{\rm ring}+0.2\kpc$. Figure \ref{fig:sfr} plots temporal changes of the star formation rate (SFR) as well as the gas mass ($M_g$) in the cold-disk (left) and warm-disk (right) models. The top, middle, and bottom panels give the SFR occurring in the ring, bar, and arm regions, respectively.

In the outer disk, star formation occurs mostly inside spiral arms and is stronger in the cold disks with stronger arms.
The presence of strong spirals before the development of a bar
is responsible for a sharp increase in the arm SFR at the expense the gas mass in the outer parts of the cold-disk models. As the spirals become weaker due to heating of star particles scattered off arms and gas clouds, the SFR in the outer disk decrease rapidly with time.
The decrease of the gas mass and no gas inflow from outside also make the arm SFR decreased.

The formation of a bar certainly triggers star formation in
the bar and central regions. As the bar grows,
the non-axisymmetric potential produces a pair of dense ridges
and a nuclear ring
in which most of the disk star formation takes place at late time.
The early increasing trend of the bar/ring SFR is similar to that of the bar strength.
After the bar achieves a full strength ($t\sim0.5\Gyr$ in the cold disks and $t\sim2\Gyr$ in the warm disks), the density of dust lanes become reduced and the bar SFR experiences a dramatic drop,
while the decrease in the ring SFR is only mild due to continued mass infalls.
This is in contrast to the cases with a fixed bar potential where
fast gas exhaustion caused by a fast bar growth on the timescale of
$\sim0.2\Gyr$ makes the ring SFR declines very rapidly afterward
(e.g., \citealt{seo13,seo14}).
The relatively slow decrease of the ring SFR
results from the fact that bars in our current models grow slowly and become longer in size over time. This not only makes the duration of star formation
extended but also expands the regions influenced by the bar potential, allowing sustained gas inflows to the rings.
In addition, mass ejections via SNe from star particles help to increase the gas mass available for the ring star formation.

Overall, the ring SFR is larger when a bar is stronger.
The gas mass, roughly  $M_{g,\rm ring}\sim 5\times 10^7\Msun$, in the ring is insensitive to $\fgas$, so that  a smaller ring has a higher gas density and thus a higher SFR.
The ring SFR is highly episodic and bursty caused by star formation feedback. Sometimes, especially at early time when rings are small,
feedback is so strong that the rings are completely destroyed and reform multiple times. Column (7) of Table \ref{tbl:model} lists the time-averaged ring SFR over $t=4.5$--$5.0\Gyr$. In model {\tt W10}, the bar is weak and grows slowly, and the resulting SFR in the ring also exhibits a slow growth with intermittent bursts.  Since the gas stays longer in the bar region, the bar SFR in model {\tt W10} is
higher than that in models {\tt W05} and {\tt W07} with stronger bars. In model {\tt C10}, the bar size remains almost unchanged after $t\sim 1\Gyr$, so that the ring SFR keeps decreasing as the bar region becomes devoid of gas.

\section{Summary and Discussion}\label{sec:sum_dis}

\subsection{Summary}\label{sec:sum}

We have presented the results of
self-consistent three-dimensional simulations of barred galaxies that
possess both stellar and gaseous disks.
Our primary goals are to understand the effects of the
gaseous component on the bar formation and to explore how nuclear rings form and evolve in galaxies where the bar properties vary self-consistently with time.
We consider radiative heating and cooling of the gas and
allow for star formation and related feedback, but
do not include magnetic fields in the present work.
We consider two sets of models, similar to the Milky Way,
which differ in the velocity anisotropy parameter $f_R$
(or, equivalently, the minimum
Toomre stability parameter $Q_{T,\rm min}$).
The models with $f_R=1$ and 1.44 have a disk with
$Q_{T,\rm min}=1.0$ and 1.2,
and are thus referred to as ``cold-disk'' or
``warm-disk'' models, respectively.
In each set, we vary the mass of gaseous disk, $\fgas=M_g/M_{\rm disk}$, in the range between 0 and 10\%, while fixing the total disk
mass to $M_{\rm disk}=5\times 10^{10}\Msun$.
The main results of our work can be summarized as follows.

\begin{itemize}
\item[1.] \emph{Effects of Gas on the Bar Formation}:
Perturbations in the initial disks are swing amplified to form spiral structures. In the cold-disk models, the initial swing amplification is strong enough to make the spirals highly nonlinear instantly, and the disks soon become dominated by the $m=3$ spirals that have a long duration for growth. These $m=3$ spirals interact nonlinearly with other modes with different $m$ and rapidly transform to an $m=2$ bar mode supported by closed $x_1$-orbit families. Since the gaseous component is effectively colder than the stellar component, a bar in a disk with larger $\fgas$ forms faster and more strongly.

In the warm-disk models, however, the initial swing amplification is only moderate and the resulting spirals are in the linear regime.  Thus, $m=2$ spirals that will eventually become a bar should be amplified further via successive swing amplifications and multiple loops of feedback. Modes with larger $m$ are favored in a disk with larger $\fgas$ due to lower effective velocity dispersions, indicating that the amplitude of the bar-seeding $m=2$ spirals is lower for larger $\fgas$. Unlike in the cold disks, therefore, a warm disk with larger $\fgas$ forms a bar more slowly.

\item[2.] \emph{Bar Evolution}:
Bar formation necessarily involves the mass re-distribution as well as gas inflows toward the center, increasing CMC.  The CMC as well as the bar mass in turns weaken the bar by exciting stellar motions in the vertical direction. The CMC grows faster for a stronger bar, resulting in a faster bar decay. For example, a bar in model {\tt C10} grows very rapidly ($\sim0.1\Gyr$) and then becomes weaker by a factor of three in $\sim1\Gyr$, eventually turning to an oval. On the other hand, model {\tt C00} with a slow increase of the CMC does not experience such bar weakening. Consequently, bars in both cold and warm disks become stronger in models with smaller $\fgas$ at the end of the runs. The bar length $R_b$ is correlated with the bar strength such that a stronger bar is usually longer.  Although bars are fast when they form, they slow down to become slow rotators with $\mathcal{R}=R_{\rm CR}/R_b > 1.4$ by transferring
angular momentum to their surrounding halos.

We find that all bars formed in our models thicken over time to become a B/P bulge.  While the bar thickening occurs gradually due to vertical heating in most models, the gas-free, warm-disk model {\tt W00} undergoes a rapid thickening via buckling instability which occurs when $\sigma_z/\sigma_R\lesssim 0.5$, where $\sigma_z$ and $\sigma_R$ refer to the velocity dispersions in the vertical and radial directions, respectively. The presence of gas tends to stabilize the buckling instability by enhancing the CMC and thus $\sigma_z$, consistent with the results of \citet{ber07} and \citet{ian15}. Although model {\tt C00} does not have gas, its CMC growth is strong enough to quench the buckling instability.

\item[3.] \emph{Nuclear Ring and Spur}:
The non-axisymmetric bar torque induces shocks in the gas flows
and form dust lanes.  The gas experiences infall along the dust lanes to form a nuclear ring. At early time when the bar grows,
only a gas close to the galaxy center responds to the bar potential, leading to a small nuclear ring with radius $\rring\sim40\pc$. Due to strong feedback from explosive
star formation inside the ring, the tiny ring is repeatedly disrupted and reforms. As the bar grows in size, gas at larger radii starts to infall and be added to the nuclear ring.
Since the gas at larger radii has increasingly larger angular momentum, the addition of gas from larger radii makes the nuclear ring larger with time, up to $\rring\sim 0.2$--$0.5\kpc$ at the end of the runs. Overall, the rings are larger in the cold disks since they form earlier and thus grow for a longer period of time.
The ring is smallest in model {\tt W10} in which the CMC offsets
the effect of the bar size on the ring growth  \citep{li17}. The ring is largest in model {\tt C10} where spiral arms supply
gas with high angular momentum to the bar regions.

Unlike the previous simulations with a fixed bar potential,
our current self-consistent simulations form filamentary interbar spurs that are connected perpendicularly to dust lanes.
The origin of filaments is expanding SN shells produced
by star formation feedback that are sheared out in the low-density bar regions. Since the bars becomes stronger and longer over time, the dust lanes move gradually away from the bar semi-major axis. When the filaments hit the dust lanes perpendicularly, the local density is enhanced by an other of magnitude in the dust lanes, sometimes enough to form stars.

\item[4.]  \emph{Star Formation}:
The cold-disk models form stars both in the outer disks with spiral arms and in the inner disk with a bar, while star formation
in the warm-disk models with weak spirals is concentrated in the inner disk. Bar formation triggers star formation in the bar regions (mostly inside
dust lanes) as well as in the nuclear rings. Overall, the ring SFR is stronger for a stronger bar. The ring star formation is highly episodic and bursty due to feedback that can sometimes disrupt the rings. The SFR in the bar regions rapidly declines after the
bar attains the peak strength.
However, the decrease in the ring SFR is quite mild due to a slow bar growth as well as a temporal increase in the bar length, the latter of which can continuously supply the gas to the ring at late time. Mass return via SNe also helps the ring SFR persist
longer than the cases with a fixed bar potential (e.g., \citealt{seo13,seo14}).
Overall, the ring SFR is very similar to the the mass
inflow rate to the ring, amounting typically to
$-0.1$ to $-0.5\Aunit$ at velocities $150$--$200\kms$ along
the dust lanes (see Figure \ref{fig:minfall}).

\end{itemize}

\subsection{Discussion}\label{sec:dis}

In this paper, we consider galaxy models similar to the Milky Way to study bar formation in disks with gas.  The properties of the bars and nuclear rings formed in our simulations are very close to those in the Milky Way. The Milky Way is known to have a bar with the semi-major axis of $R_b\sim 3$--$5\kpc$  (\citealt{mor96,dam01,fer07,kru15,bla16}).
The CMZ, a nuclear ring in the Milky Way, has a radius of $\rring\sim0.2$--$0.5\kpc$ \citep{mor96}, consistent with the ring sizes displayed in Figure \ref{fig:ring_size}. The CMZ is observed to have a total gas mass of $\sim 3$--$7\times10^7\Msun$ \citep{dah98,fer07,imm12,lon13}, similar to gas mass  $M_{g}\sim3$--$5\times10^7\Msun$ in the ring of model {\tt W05}.
The present-day SFR in the CMZ is estimated to be $\sim0.04$--$0.1\Aunit$(\citealt{mor96,yus09, imm12, lon13, koe15}), about 10 times smaller than the value inferred from the CMZ mass
\citep{tsu99, lon13}. This is probably because star formation in the CMZ is episodic and currently in the low state \citep{kru14,kru16},
consistent with the results of our simulations.

\begin{figure}
\epsscale{1.1} \plotone{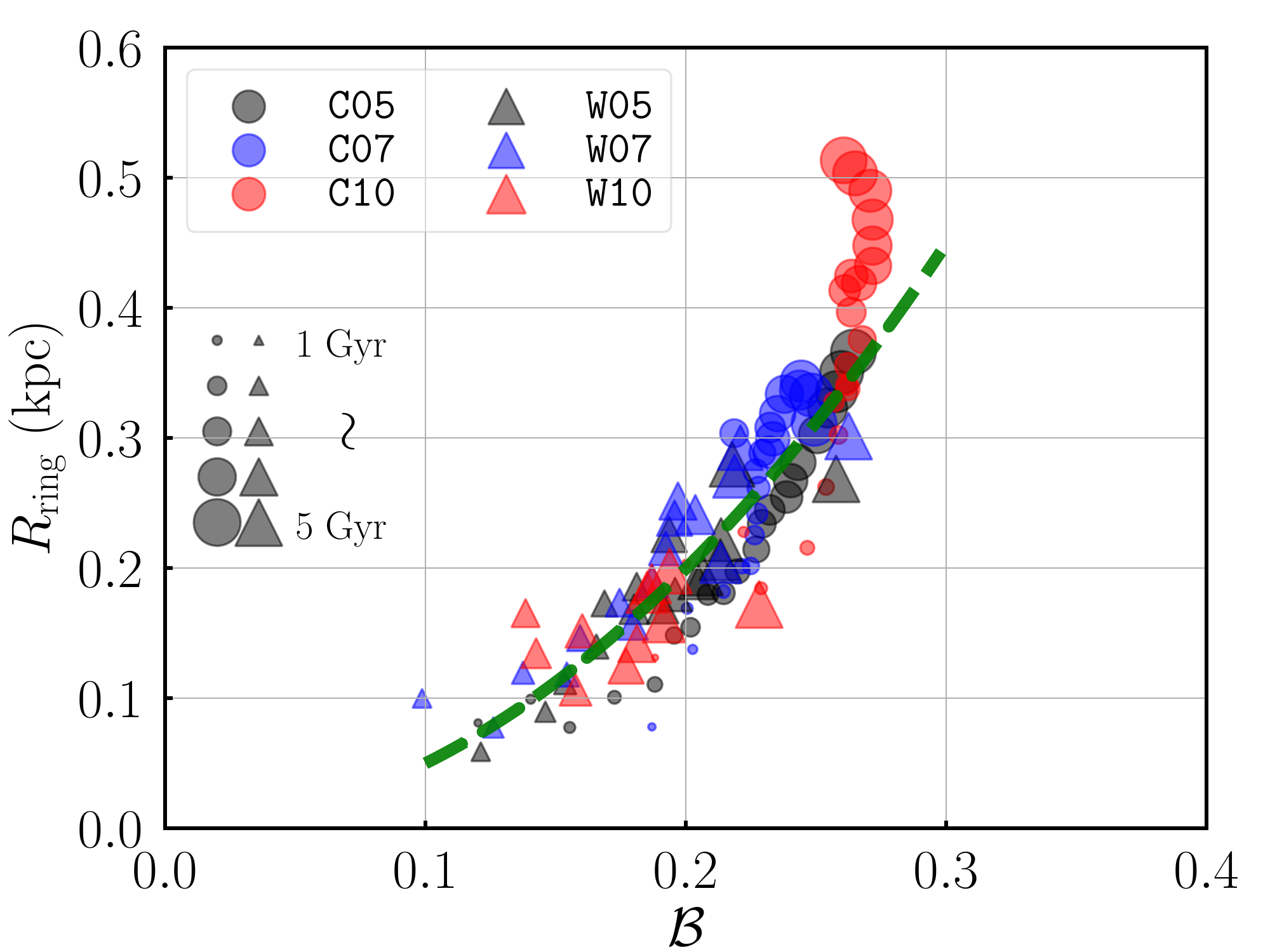}
\caption{Sizes of nuclear rings against the dimensionless parameter
${\cal B} \propto {\rm CMC}/(A_2^{0.3} \Omega_b^{0.2})$ defined in Equation \eqref{e:Bparam}. The size of symbols corresponds to the simulation time. The dashed line draws  $R_{\rm ring}/(1\kpc)=5{\cal B}^2$, which fits the numerical results reasonably well, except model {\tt C10}.
\label{fig:RringB}}
\end{figure}

Our models show that the size of a nuclear ring increases with time after the temporal bar weakening associated with the CMC, suggesting the CMC definitely affects the ring size.  Previous simulations
employing a rigidly-rotating fixed bar potential found that rings
are smaller for stronger and/or faster bars \citep{kim12a,li15}.
To see what controls the ring size $R_{\rm ring}$ in our current models in which bars evolve self-consistently, we try to fit $R_{\rm ring}$ using various combinations of the CMC, the bar strength $A_2$, and the pattern speed $\Omega_b$, and empirically find that the dimensionless parameter
\begin{equation}\label{e:Bparam}
  {\cal B} \equiv A_2^{-0.3} \left(\frac{{\rm CMC}}{10^{10}\Msun}\right)
  \left(\frac{\Omega_b}{1\kms\kpc^{-1}}\right)^{-0.2}
\end{equation}
gives a reasonable fit.
Figure \ref{fig:RringB} plots $R_{\rm ring}$ as a function of ${\cal B}$ for all models at various time, with the symbol size denoting the time. Except for model {\tt C10} with an oval, the ring size is given
roughly by $R_{\rm ring}/(1\kpc) = 5{\cal B}^2$,
showing that nuclear rings are larger for larger CMC, smaller $A_2$,
and/or smaller $\Omega_b$, consistent with the results of the previous simulations (e.g., \citealt{kim12a,li15}) and
observations (e.g., \citealt{maz08,maz11,com10}).
The dependence of $\cal B$ on the physical parameters in
Equation \eqref{e:Bparam} suggests that the ring size
in our simulations is most sensitive to the CMC rather than
the bar strength or pattern speed.

Except model {\tt C10} with an oval, all bars in our models have
${\cal R}=R_{\rm CR}/R_b\sim 1.5$--$1.8$ during most of their evolution, corresponding to slow bars. This is in contrast to observational results that bars in most external galaxies are fast rotators with $1<{\cal R} <1.4$ (e.g., \citealt{cor08,cor11,fat09,per12,agu15}). A recent made-to-measure modeling of \citet{por17} to match the red clump giant density as well as the bulge kinematics obtained from various surveys found
$\Omega_b=39 \kms\rm\;kpc^{-1}$ for the bar in the Milky Way,
with the corotation resonance $R_{\rm CO}=6.1\kpc$. Together with the
measured bar length of $R_b=5.0\kpc$ \citep{weg15}, this suggests that the Milky Way bar is also a fast rotator with ${\cal R}=1.22$ \citep{por17}. It appears that the bar pattern speed as well as ${\cal R}$ are affected by various parameters such galaxy rotation curve, gas fraction, halo shape, etc.\ (e.g., \citealt{ath14,pet18}). For instance, \citet{pet18} showed that the bar pattern speed depends rather critically on the shape of the rotation curve in such a way that bars under the ``rising" rotation curve are slow,
while the other rotation curves produce fast bars.
We note that the our rotation curve shown in Figure \ref{fig:tq} is quite similar to the rising rotation curve considered in \citet{pet18}. Under the rising rotation curve, the bars that form have relatively low specific angular momentum and thus a small pattern speed from the outset, and a small amount of angular momentum transfer to the halo makes them evolve to slow bars. An inclusion of a strong bulge, which is missing in our current models, would make the bars rotate faster.

All nuclear rings formed in our models have radii
less than $\sim0.6\kpc$ at the end of the runs. Although these
are more-or-less comparable to the ring sizes in galaxies like the Milky Way, they are certainly smaller than typical nuclear rings in normal barred-spiral galaxies such as NGC 1097 (e.g., \citealt{com10}) and those formed in simulations with fixed bar potentials (e.g., \citealt{kim12a,kim12b,li15}). Relatively small nuclear rings are presumably due to the absence of a bulge in our initial galaxy models. Recently, \citet{li17} used hydrodynamic simulations with static stellar potentials to show that a nuclear ring forms only in models with a central object exceeding $\sim1\%$ of the total disk mass, and that the ring size increases almost linearly with
the mass of the central object.
This opens the possibility that the presence of a massive compact bulge would make a ring large when it first forms.  The ring can be even larger as it grows due to an addition of gas with larger angular momentum from outer regions.

We find that the effects of gas to the bar formation depends rather sensitively on the velocity anisotropy parameter or $Q_{T,\rm min}$ such that gas causes a bar
to form faster and stronger in cold disks with $Q_{T,\rm min}=1.0$, while tending to suppress the bar formation in warm disks with $Q_{T,\rm min}=1.2$.
In contrast, \citet{ath13} reported that the gaseous component with $\fgas\leq 50\%$ always prevents the bar formation, similarly to our warm-disk models. On the other hand,
\citet{rob17} ran simulations of bar formation with or without AGN feedback, and showed that gas in models with AGN feedback promotes
the bar formation, similarly to our cold-disk models, whereas the bar
formation without AGN feedback is independent of $\fgas$.
These discrepancies in the results of various simulations with different parameters suggest that bar formation involves highly nonlinear processes especially with gas and is thus very sensitive to the initial galaxy models as well as the gas fraction.

In addition to the gas fraction and velocity anisotropy parameter,
the properties of a DM halo also appears to affect dynamical
evolution of bars via angular momentum exchanges with the disks
(e.g. \citealt{sel80,deb00,ath03}).
Recent numerical simulations (with no gas) showed that the bar evolution is influenced by the shape and the spin parameter of the DM halo (e.g., \citealt{ath13,col18}).
In particular, \citet{ath13} found that bars under a triaxial halo form earlier and experience stronger decay than in galaxies
with a spherical halo. On the other hand, \citet{col18} found that bars under both prolate and oblate haloes start to form later. They further showed that a halo with faster spin is less efficient in absorbing angular momentum and thus results in a weaker and smaller bar. It will be interesting to see how the presence of gas conspires with the halo spin to guide the bar formation and ensuing evolution.

In this work, we did not consider feedback from the central black hole that was allowed to accrete the surrounding gas passively.
In the Milky Way, the observed fluorescent X-ray emission from cold
iron atoms in molecular clouds inside the CMZ is possibly due to
X-ray irradiation from Sgr A*, suggesting potential importance
of the AGN feedback on nuclear rings (e.g., \citealt{koy96,su10}).
Some barred galaxies host AGN at their centers, but the
physical connection between a bar and AGN activities is not clear.
Some observational studies suggest that the bar fraction in AGN-host galaxies is higher than in galaxies without AGN (e.g., \citealt{ars89,lai02,gal15}),
while other studies do not find any specific correlations between them (e.g., \citealt{ban09,lee12b,che14}).
Recently, \citet{rob17} used numerical simulations to find
that AGN feedback suppresses star formation in the vicinity of a
black hole, while forming a dense ring in which star formation
is enhanced.  They found that such positive and negative effects
are almost equal, making no overall
quenching or enhancement of star formation in barred galaxies.
We note that these results were based on models with a static (rather
than live) halo and a not-well-resolved gas disk with $N_g=1.2\times10^5$ particles.  It is desirable to run self-consistent models with high resolution to accurately assess the effects of AGN feedback on star formation in barred galaxies.

\acknowledgments

We gratefully acknowledge a thoughtful report from the referee,
and helpful discussions with Eve Ostriker.
This work was supported by the grant (2017R1A4A1015178) of National
Research Foundation of Korea.
The computation of this work was supported by the Supercomputing
Center/Korea Institute of Science and Technology
Information with supercomputing resources including technical
support (KSC-2018-C3-0015).

\textit{Software:} GIZMO(\citealt{hop15}), GalIC(\citealt{yu14}),
additional data analyses and visualizations were done using IDL version 8.6 and IPython (\citealt{per07}).

\end{document}